# Understanding early stages of low-temperature hydrogen-driven direct co-reduction of Fe-Ni mixed oxide thin films at the near atomic scale

Emmanuel Uwayezu [a], Shaolou Wei [b]*, Yujiao Li [c]*, Johannes D. Bartl [a], Dierk Raabe [b]*, Alfred Ludwig [a c]*

[a] Chair for Materials Discovery and Interfaces, Institute for Materials, Ruhr-Universität Bochum, Bochum, Germany

[b] Max Planck Institute for Sustainable Materials, Düsseldorf, Germany

[c] Center for Interface Dominated High Performance Materials (ZGH), Ruhr-Universität Bochum, Bochum, Germany

*Corresponding authors*

Email: sl.wei@mpie.de, yujiao.li@rub.de, d.raabe@mpie.de, alfred.ludwig@rub.de

## Abstract

Kinetic understanding of the hydrogen co-reduction of multinary and multi-phase oxides is of interest for enhancing sustainability of alloy production and transition to a hydrogen-based economy. Benefits include decrease in energy consumption, enhanced kinetics, conversion of oxides to alloys in a single step, and opportunities to create microstructures that are hard to realize by other processes. Thin films provide a versatile platform to study these processes as reactive co-deposition from multiple elemental, alloy or compound targets and precise oxygen flow control allow atomic mixing into various oxide phases which are well-defined nanoscale precursor structures for the subsequent reduction study at the near atomic scale. Using this approach, the early stages of the hydrogen direct reduction of oxide thin films are investigated using a $Fe_{50}Ni_{50}O_x$ thin film consisting of $NiFe_2O_4$ and $NiO$ phases as a case study. After reduction at 280°C in pure $H_2$ for different times, structural, morphological, and nanoscale changes were examined by different characterisation methods including atom probe tomography (APT). It was found that the film's low-temperature reduction is nucleation-limited marked by grain-boundary nucleation preceded by an incubation time of more than 5 min. APT revealed that the early-stages of the reduction involves phase separation into a Ni-rich $Fe_xNi_y$ metallic phase and a transformed remaining oxide (magnetite, $Fe_3O_4$). Further reduction induces magnetite reduction and alloying into a nearly equiatomic FeNi alloy. The low-temperature reduction and alloying are facilitated by synergetic effects from the film's nanostructure, and Ni autocatalytic effects through alloying and hydrogen spillover. The results pave the way for low-temperature formation of Fe-Ni alloy thin films with tunable compositions directly from oxides, and broaden the scope of hydrogen direct reduction of multinary oxides to thin-film platforms.

# 1 Introduction

Hydrogen direct co-reduction of multicomponent oxides offers an attractive method for sustainable alloy synthesis directly from oxides [1–3]. Binary [1,4] and complex [2] bulk alloys with tunable composition and properties have been synthesised in an integrated single step from mechanically and chemically mixed oxides resulting in substantial energy savings with near-zero greenhouse gas emissions [1,2]. It also offers additional benefits such as lower reduction temperatures [4], faster kinetics and autocatalytic effects through hydrogen spillover [5,6]. Despite these benefits, complex convolution of thermodynamic and kinetic contributions from individual metal oxide precursors during reduction, particularly in compositionally complex alloys, makes systematic understanding of the underlying mechanisms and optimisation of the process challenging.

This complexity renders thin films compelling as platforms for studying the hydrogen reduction of multinary and multi-phase oxides: thin film deposition processes including sputtering enable precise control over the chemistry [7,8], geometry [9,10], and structure [11,12] of the films which can facilitate the design of exact model systems for accurate observation of the transformation allowing the decoupling of thermodynamic and kinetic effects to infer the underlying rate determining steps and unravel intermediate phase transitions. For instance, Patterer et al. [13] made dense Si-doped $Fe_2O_3$ thin films using reactive high power pulsed magnetron sputtering to study the chemical effects of Si on hydrogen direct reduction (HyDR) of the Fe oxide — avoiding interference from structural porosity. By annealing the doped oxide film in an Ar+5%$H_2$ atmosphere at 700°C, they found that Si diffuses out of the film towards the surface and reacts with oxygen forming a $SiO_x$-rich near-surface layer, blocking the available surface O sites – a purely chemical effect independent of any effects from porosity [13]. A Si content of as low as 3.7 at.% was enough to effectively inhibit the reduction of the oxide film.

In addition, with thin films, various oxide systems ranging from binary over multinary to complex mixtures can also be realised using physical vapour deposition techniques such as reactive sputtering as they allow co-deposition from multiple elemental [14] or compound targets [15] and fine-tuning of the $O_2$ flow which results in elemental atomic mixing into oxide phases that can be difficult to achieve through other means [16]. Multilayers [17], epitaxial films [18], and thin-film material libraries covering binary [19] or multinary [14–16] composition spaces can be synthesised.

Fundamentally, the reduction of transition metal oxides is governed by the active kinetic coupling amongst surface reactions [20], mass transport processes [21,22], mass loss [21], and charged defect evolution within the oxide lattices [22]. In thin-film geometries, a pronounced surface-to-volume ratio [23] and significantly reduced diffusion length scales [24] often amplify these kinetic mechanisms, enabling further the decoupling of thermodynamic reducibility from kinetic limitations, but also leading to reduction pathways that can deviate from their bulk counterparts [25]. Still, considering that mean square diffusion distances in bulk oxides at temperatures (typically 500-1100°C [2,26,27]) and timescales (typically an hour or less [3,4,28]) used for HyDR falls in the range [29,30] of thin film dimensions (from nanometres to a few micrometres in thickness [31]), this would provide key insights relevant for the understanding and optimisation of bulk hydrogen direct co-reduction processes. Therefore, thorough understanding of these kinetic contributions in transition metal oxide thin films is needed for rational alloy design and for unravelling hydrogen reduction mechanisms under conditions that are far from bulk equilibrium, where geometric confinement, strain, and interface effects can play an eminent role.

In this study, we investigated early stages of the low-temperature hydrogen-based direct co-reduction of Fe-Ni-O thin films by using a $Fe_{50}Ni_{50}O_x$ thin film, deposited by reactive co-sputtering [32,33], as a case study. The as-deposited film — a two-phase mixture of $NiFe_2O_4$ and $NiO$, was reduced at 280°C for different times using pure $H_2$ and changes in the structure, morphology, and nanoscale chemistry were examined using different characterisation methods including X-ray diffraction (XRD), scanning electron microscopy (SEM) and atom probe tomography (APT). The vital role of grain boundary nucleation at such temperatures, and faster reduction kinetics due to the nanostructure of the film were elucidated. Fe-Ni is an important alloy system whose low temperature phase diagram remains largely unexplored due to slow Fe-Ni interdiffusion at temperatures below about 400°C [34]. The findings provide a basis for sustainable synthesis of Fe-Ni alloy thin films with tunable compositions directly from oxides, and could be useful in the design and optimisation of bulk HyDR processes.

## 2 Experimental details

### 2.1 Material synthesis

An Fe-Ni oxide thin film was synthesised by reactive co-sputtering from 100 mm diameter elemental targets positioned confocally 180° apart in a magnetron sputter system (ATC-2200 V, AJA International, U.S.A.). Prior to deposition, a 5 min process stabilisation step was conducted with closed shutters using the deposition parameters described below. The film was deposited on a rotating $Si/SiO_2$ (500 nm thermal $SiO_2$) substrate at 800 K preheated for 1 h. The base pressure in the sputter chamber was $5.3 \times 10^{-6}$ Pa at room temperature (RT) and $2.4 \times 10^{-5}$ Pa after preheating. Deposition was conducted at a pressure of 0.4 Pa and gas flow rates of 20 sccm and 40 sccm for $O_2$ and Ar respectively. A 210 W pulsed direct current (pulsed DC) power (350 kHz, 1.4 μs pulse length) was applied to the Fe target (purity 99.990%) and a 210 W radio frequency (RF) power was applied to the Ni (purity 99.995%) target for 145 min achieving an oxide film with a nominal thickness of 100 nm (based on pre-determined sputter rates) and an equiatomic ($Fe_{50}Ni_{50}O_x$) metal composition, i.e. 50 at.% Fe and 50 at.% Ni (normalised to the content of metallic elements).

## 2.2 Solid state reduction

Reduction was conducted in a tubular quartz glass reactor (36 mm diameter) inside a horizontal three-zone high temperature furnace, equipped with a gas supply system. The sample was placed in a quartz glass boat and inserted into the reactor which was then purged with He (flow rate = 6 L h$^{-1}$) for 45 min at RT. The reactor was then heated under He with a flow rate of 10 L h$^{-1}$ and heating rate of 5 K min$^{-1}$ up to a temperature set point of 280°C. The sample was then reduced isothermally at 280°C with pure H$_2$ (99.999%) for different times (5, 7, and 10 min). The H$_2$ flow rate was 10 L h$^{-1}$. After reduction, the sample was furnace cooled under He with a flow rate of 10 L h$^{-1}$ down to 100°C followed by 3 L h$^{-1}$ He flow for the remainder of the cooling time.

## 2.3 Characterisation

The thin film oxide was characterised both before (as-deposited) and after reduction to investigate reduction-induced changes. Microstructural and compositional examinations were done using scanning electron microscopy (SEM) and energy dispersive X-ray spectroscopy (EDS) in a JEOL JSM-7200F SEM with an Oxford AZtecEnergy X-MaxN 80 mm$^2$ silicon drift EDS detector. Crystallographic phase constitution in the different states was studied using X-ray diffraction (XRD) in a Bruker D8 Discover diffractometer fitted with a Vantec-500 2D detector and a Cu Kα source in Bragg–Brentano geometry. A 1 mm collimator was used for the measurements and 2D-images (frames) were collected at five 2θ positions from 20° to 80° in 15° steps. An offset of 2.5° was used on θ to circumvent major Si (100) substrate peaks. The Inorganic Crystal Structure Database (ICSD) was used for identification of the acquired diffractograms. Raman spectroscopy was used to complement XRD to elucidate the nature of the spinel phase that forms in the thin film oxide. Measurements were performed at RT with a Renishaw InVia microscope using a green laser (λ = 532 nm) with a laser power ≤ 2.5 mW, an exposure time of 60 s, and a magnification of 20x, providing a spectral resolution of 0.3 cm$^{-1}$ and lateral resolution of 0.25 µm.

Near atomic-scale investigations were conducted using atom probe tomography (APT). Needle-shaped APT specimens were prepared in a dual-beam focused ion beam (FIB/SEM) system (FEI Helios G4 CX) following a similar protocol as described by Thompson et al. [35]. Because the film is nanometric in thickness (100 nm), a 0.6 µm-thick Pt layer was deposited first to protect the region of interest (0.1 µm with the electron beam at 2 kV and 2.8 nA followed by 0.5 µm with the ion beam at 30 kV and 0.23 nA) using the system's gas injection system to minimize ion beam damage and facilitate subsequent annular milling.

APT data were collected using CAMECA LEAP 5000 XR atom probe system in laser-pulsing mode. For specimens prepared from the as-deposited state, a laser pulse energy (LPE) of 10 - 60pJ, a pulse repetition rate of 200 kHz, and a detection rate of 5 ions per 1000 pulses were used. For the reduced (5 and 7 min) specimens (and one as-deposited specimen for direct comparison), a LPE of 50 or 60 pJ, a pulse repetition rate of 125 kHz, and a detection rate of 3.5 ions per 1000 pulses were selected to reduce premature fracture, which results from high nanoporosity in the film upon reduction. For both cases, a base temperature of 60 K was used. Reconstruction and data processing were performed using CAMECA's APsuite 6.3 software. The fixed angle reconstruction protocol was used, and initial tip diameter ($d_0$) and shank angle ($\theta_0$) were measured in SEM for each ready APT specimen.

## 3 Results

### 3.1 Phase constitution in the as-deposited and reduced thin film

XRD patterns of the thin film acquired before and after isothermal reduction at 280°C for 5, 7 and 10 min are shown in **Figure 1a.** The as-deposited film consists of two phases: spinel ($NiFe_2O_4$) and nickel monoxide (NiO). No detectable phase changes are seen after 5 min reduction. However, after 7 min reduction time, phase decomposition starts: a new phase emerges, as indicated by the broadening and shifting of one of the peaks highlighted by a vertical dashed line in **Figure 1a**. The dependence of transformation on isothermal holding time suggests that nucleation plays a critical role in the process. The reduction product must overcome a substantial nucleation barrier, which explains the observed incubation time particularly at these low temperatures where the thermodynamic driving force is modest.

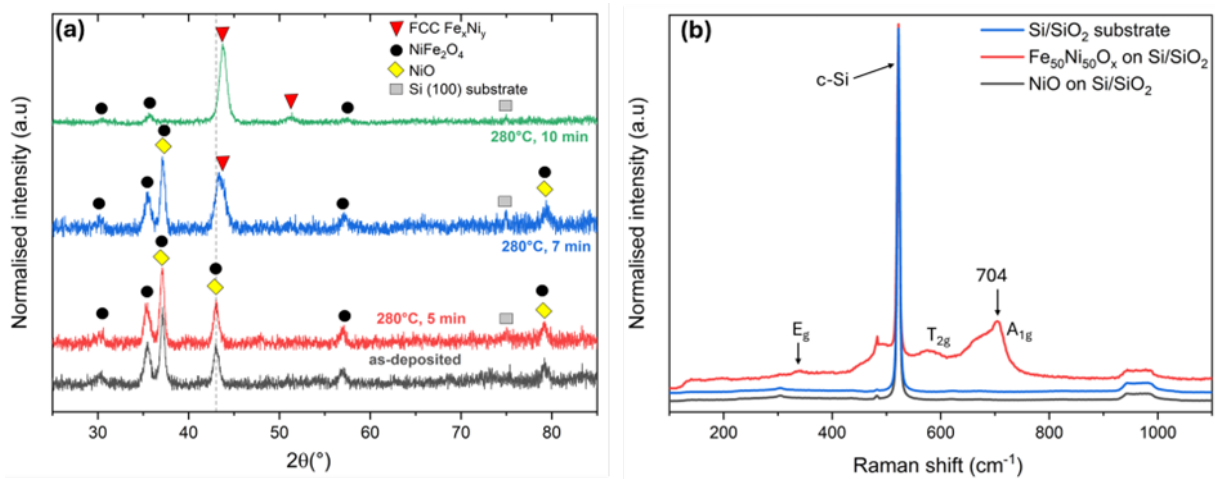

**Figure 1**. a) XRD analysis of the $Fe_{50}Ni_{50}O_x$ thin film before and after reduction at 280°C for different times. The as-deposited state is shown in black. The reduced states are presented in red, blue, and green, corresponding to reduction at 280°C for 5 min, 7 min, and 10 min, respectively. The peaks corresponding to FCC $Fe_xNi_y$ (red triangle), $NiFe_2O_4$ (black circle), NiO (yellow diamond), and the Si (100) substrate (gray square) are indicated. b) Raman spectrum of the as-deposited film. NiO and $Si/SiO_2$ substrate spectra are included for comparison. All spectra were normalised to the crystalline Si (c-Si) peak at 522 $cm^{-1}$.

Ni ferrite ($NiFe_2O_4$) and magnetite ($Fe_3O_4$) crystallise in the same (inverse) spinel structure [36,37]. Although $NiFe_2O_4$ is reported to exhibit a slightly smaller lattice parameter than $Fe_3O_4$, i.e., 0.83385(3) nm [38] versus 0.83967(3) nm [39], such a subtle difference is difficult to unambiguously capture by the lab-scale XRD, especially in the presence of potential residual strain in the thin films [40–42]. Therefore, to identify the nature of the spinel phase, Raman spectroscopy was used, and the results are presented **in Figure 1b.** Spectra of a NiO thin film deposited with the same parameters as the $Fe_{50}Ni_{50}O_x$ film, as well as an uncoated $Si/SiO_2$ substrate are included as references to identify potential peak overlaps (**Figure 1b**). The most intense Raman peak for the as-deposited film is the $A_{1g}$ band which appears at 704 cm$^{-1}$ with a shoulder at the lower wave number side of the band that is characteristic for the $NiFe_2O_4$ spinel [43]. Normally, five ($A_{1g}$, $E_g$, and 3 $T_{2g}$) bands are Raman active for spinels [44]. The shoulder or doublet bands observed in the $NiFe_2O_4$ Raman bands arise from short-range 1:1 ordering of $Fe^{3+}$ and $Ni^{2+}$ at the octahedral sites. This ordering effect lowers local symmetry and produces non-uniform Fe-O and Ni-O bond lengths [45]. The non-uniformity has also been attributed to distortion originating from differences in charge and ionic radii of $Fe^{3+}$ and $Ni^{2+}$ [43]. In addition to the $A_{1g}$ band, a broad $E_g$ band is also present in the Raman spectrum of the as-deposited film at 331-243 cm$^{-1}$. Only one $T_{2g}$ band is visible due to overlaps with substrate bands. This finding is consistent with those of other studies on bulk, thin film, and nanostructures of $NiFe_2O_4$ [43–46], confirming that the as-deposited film is a two-phase mixture of $NiFe_2O_4$ + NiO, also in agreement with a previous report on the equiatomic Fe-Ni oxide system [47]. **Figure 1b** shows that NiO does not exhibit first order Raman bands under ideal symmetry. As mentioned by Dietz et al. [48], first-order Raman bands are forbidden by selection rules for the oxide due to crystallographic inversion symmetry of the rock salt crystal structure, and are only present when the symmetry is broken by a high concentration of Ni vacancies [49–51].

The volume fraction of the reduction product phase increases with increasing the holding time further to 10 min (indicated by the increase in its relative intensity and the appearance of an additional peak), and this phase is identified as an FCC (face - centred cubic) - like metallic phase, as shown in **Figure 1a**. In addition, all peaks related to NiO are no longer visible after 10 min of reduction, suggesting that NiO is the first phase to be reduced, and the new phase is expected to be Ni-rich, as proposed in several studies on the bulk mixed Fe-Ni oxide system [1,27].

### 3.2 Microstructural evolution of the thin film reduced in hydrogen at 280°C

Surface morphological changes due to reduction were investigated with SEM and the results are shown in **Figure 2**. The as-deposited film consists of a relatively smooth and uniform microstructure (**Figure 2a&b**), with nanostructured grains as shown by a high resolution SEM image (see **Figure S1** in the supplementary materials). No apparent morphological changes were observed after 5 min reduction at 280°C (**Figure 2c**), consistent with XRD results. Significant microstructural changes appear after reduction at 280°C for 7 min (**Figure 2d-f**) marked by the formation of domains of a new phase (reduction product), with a spherical morphology, within the unreduced/remaining oxide matrix. EDS maps in **Figure 2g** show that the new phase is indeed oxygen deficient (shown by yellow arrows), in line with XRD results, corresponding to the appearance of the metallic FCC-like $Fe_xNi_y$ phase.

Neighbouring individual spherical domains of the new phase agglomerate and then coalesce into semi-continuous structures driven by surface and interfacial energy minimisation (**Figure 2e**). In some parts of the microstructure, intense coarsening occurs leading to the formation of large islands of up to 81×38 µm$^2$ in size (see **Figure S2** in supplementary materials). The evident microstructural heterogeneity is a common feature of solid-state hydrogen reduction of metallic oxides, stemming from non-uniform local thermodynamic and kinetic environments [21]. Nucleation and growth of the reduction-product phase alter the local environment, giving rise to subsequent nucleation in other parts of the microstructure (**Figure 2e&f**). **Figure 2f** also shows that the new/reduced phase is highly porous (highlighted by white arrows), which is another characteristic feature of solid-state reduction emanating from coalescence of O vacancies into nano-micropores in the reduced regions [52]. The volume fraction of the pores could not be estimated based on SEM analysis because most of the pores were < 50 nm in size. After 10 min exposure time to hydrogen, the unreduced oxide matrix is no longer visible (**Figure 2h**) indicating that more extensive reduction and coarsening take place resulting in a highly porous structure with pockets of different (darker) contrast, presumably corresponding to partially reduced oxide regions (see **Figure S3** in the supplementary materials).

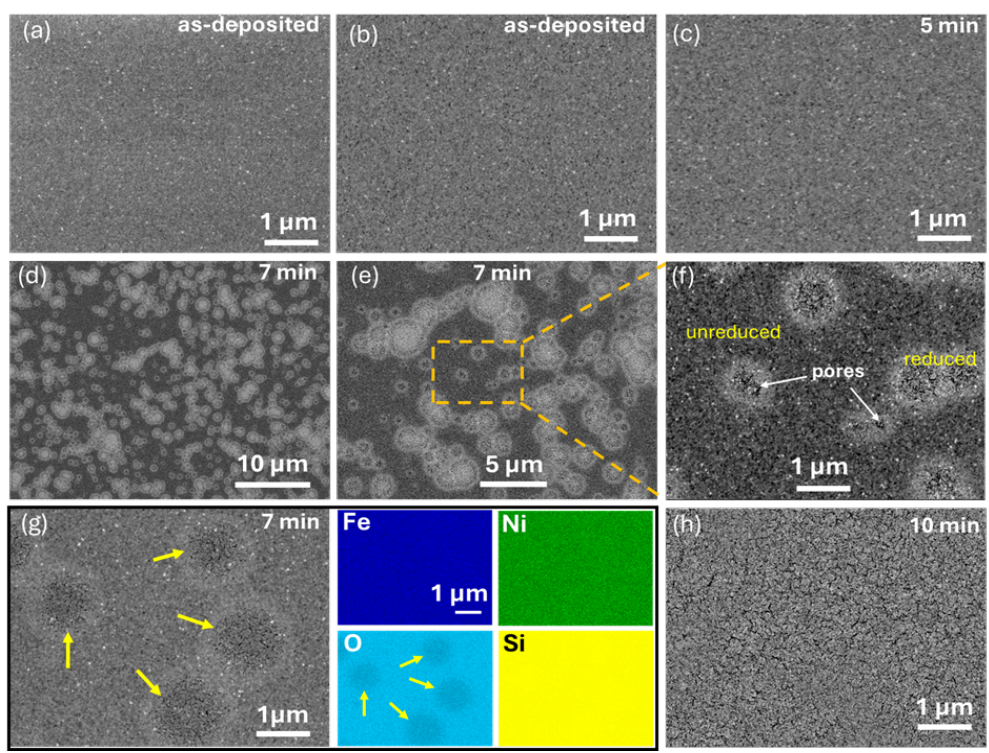

*Figure 2*. SEM analysis of Fe$_{50}$Ni$_{50}$O$_x$ thin film in different states: as-deposited and after reduction at 280°C for 5, 7, and 10 min. a) secondary electron (SE) image and b) backscattered electron (BSE) image of the as-deposited film. c) BSE image of the film after 5 min reduction at 280°C. d-f) BSE images after 7 min reduction at 280°C showing spherical domains of a new phase (reduction product) nucleating within the oxide matrix. g) BSE image and corresponding EDS map showing that the new phase is indeed oxygen deficient. h) BSE image showing the microstructure of the film after 10 min reduction at 280°C.

## 3.3 Near-atomic scale characterisation of early stages of the reduction

Nanoscale transformation of the oxide thin film upon hydrogen reduction at 280°C was investigated using APT. The analysis began with the as-deposited state as depicted in **Figure 3a-c**. Ion maps of Fe, Ni and O, as well as $O_2$ from a reconstruction of a specimen analysed at an LPE of 50 pJ and a pulse frequency of 125 kHz (**Figure 3a**) denote homogenous elemental mixing in the as-deposited thin film. The 3D reconstruction can be found in supplementary materials (**Figure S4a**). Note that, for convenience and visual clarity, other molecular ions apart from $O_2$ are not displayed in **Figure 3a** and in subsequent figures as a considerable number of oxygen ions evaporate as molecular oxygen ($O_2^+$) in this system, particularly at higher LPEs (>10 pJ, **Figure 4**). A one-dimensional (1D) composition profile through the reconstruction along the z-axis (evaporation direction, **Figure S4a**) also shows a consistent composition across the film's thickness with equiatomic Fe and Ni contents of about 26 at.% each and an O content of about 47 at.% (**Figure 3b**). This is identical to the volume composition of the as-deposited film as shown by **Table S1** in supplementary materials. We observed trace amounts of Cu (< 0.2 at.%), as shown in **Table S1**. This impurity likely originated from the cathode assembly that holds the Ni target. The target was sputtered through during deposition, which was only realised after APT analysis since the Cu content was below the detection limit of EDS (about 1 at.%). However, owing to the very low content, it does not play a major role in the subsequent reduction processes.

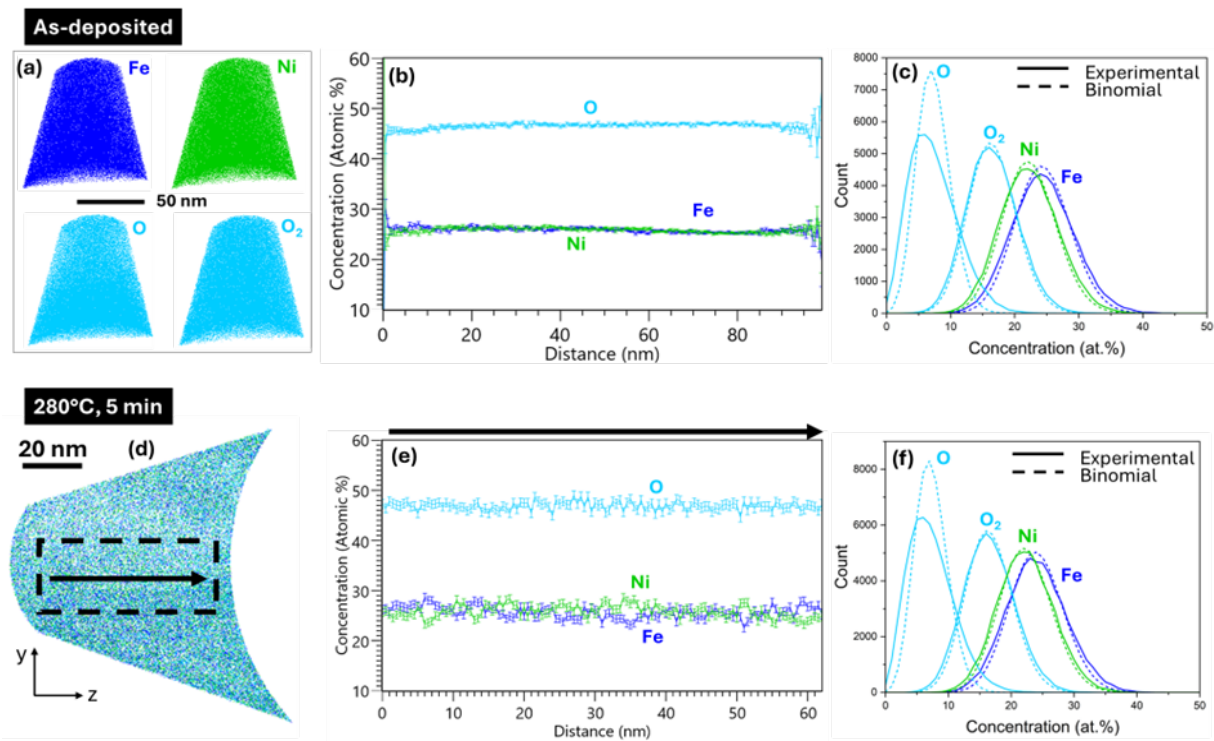

*Figure 3*. APT analysis of the as-deposited $Fe_{50}Ni_{50}O_x$ thin film (a-c): a) ion maps, b) 1D concentration profile along Z, and c) frequency distribution analysis for Fe, Ni, O, and $O_2$ ions highlighting atom-scale mixing of the two phases. APT analysis of the thin film after 5 min reduction at 280°C (d-f): d) a 10 nm slice of the reconstruction, and e) concentration profile through a 10×24.9×62.3 $nm^3$ volume of the slice showing no noticeable compositional changes at the nanoscale. f) frequency distribution analysis for Fe, Ni, O, and $O_2$ ions.

**Figure 3b** and **Table S1** also indicate that the realised elemental concentrations in the as-deposited film are different from stoichiometric predictions, with a lower O/metal ratio signifying oxygen deficit; i.e. the O concentration is below the stoichiometric composition of the different phases (57.1 at.% in NiFe$_2$O$_4$, 50 at.% in NiO, and 55.5 at.% for a NiO + NiFe$_2$O$_4$ mixture with an overall 1:1 Fe/Ni ratio) of the film (**Figure 1**). In fact, several APT specimens from the as-deposited film were analysed at various other LPEs (10, 20, 30, 40, and 60 pJ) using a pulse frequency of 200 kHz, and in each case, the O concentration was always below the stoichiometric value, i.e. underestimated, with the O/(Fe+Ni) ratio decreasing with increasing laser energy as shown by **Figure S4b** in the supplementary materials. A similar observation was made by Kim et al. [53] in their study on Fe oxides (Fe$_2$O$_3$, Fe$_3$O$_4$ and FeO). By systematically applying various combinations of laser energy (10–80 pJ), pulse frequency (65 kHz–500 kHz), and detection rate (0.5%–4%), they found that the highest O concentration achievable was still underestimated by about 5 at.% in all these oxides. This was attributed to preferential O loss during field evaporation resulting from dissociation of molecular ions into neutral oxygen, a prominent occurrence in APT analysis of ionically and covalently bonded materials including oxides [54–57], as well as nitrides [58,59]. O loss could also occur via multiple events. That is, when a single pulse triggers evaporation of multiple ions. Events that are too close in time and space can result in under-detection of certain species leading to compositional inaccuracies [60]. During the analysis of the as-deposited film, multiple events were above 40% under all the LPEs studied, decreasing from 44.81% at 10 pJ to 41.5% at 60 pJ. Similar trends and values were reported in [60] and [61]. Formation of multiple events is closely linked to molecular ion dissociation in oxides and thus provides an indirect measure of the tendency for molecular evaporation and dissociation [62]. Direct evidence of dissociation (dissociation tracks) is possible via correlation histograms first introduced by Saxey [63], but due to energy compensation, the tracks are usually not visible on data collected with a reflectron-fitted instrument [63,64] like the one used in this study. Nonetheless, Gault et al. [65] maintain that dissociation into neutral oxygen has the largest contribution to preferential O loss, and based on the high multiplicity, it is reasonable to reach the same conclusion. At lower LPEs, higher fields increase the likelihood of re-ionisation of the neutral fragments for dissociations occurring near the surface of the tip [65]. Furthermore, some field-driven dissociation [64] into charged fragments could also take place which would also be enhanced at higher fields. Therefore, more O loss is observed at higher LPEs (**Figure S4b**) despite the lower multiplicity. The peak at 16 Da was ranged as O$^+$ and potential oxygen losses from the overlap with O$_2^{2+}$ ion was not considered as no contribution from this (O$_2^{2+}$) ion at that peak was found by Bachhav et al. [66] for the Fe oxide α-hematite and Kinno et al. [67] for SiO$_2$.

**Figure 3d-f** shows the APT analysis of the film after reduction at 280°C for 5 min. The data were collected at an LPE of 50 pJ and a pulse frequency of 125 kHz, i.e., same parameters as those used for the as-deposited specimen presented in **Figure 3a-c**. No evident changes were observed, even at the nanoscale, reaffirming the existence of an incubation time at this temperature (and other reduction parameters used), in agreement with XRD and SEM observations.

Since the film consists of two different phases, as revealed by XRD analysis (**Figure 1**), the apparent nanoscale homogeneous elemental mixing was not anticipated, particularly for Fe which is only found in the spinel phase. To gain more insights, frequency distribution analysis, which compares the experimental frequency distribution of each ion in the reconstruction to the expected random (binomial) distribution was performed (**Figure 3c&f**). The Pearson coefficient, µ, provides a sample size-independent statistical measure of segregation or clustering for a given APT dataset, with values ranging from 0 (randomness) to 1 (complete clustering/segregation) [68]. Evidently (**Figure 3c**), experimental distributions of Fe (µ = 0.21), Ni (µ = 0.14), and $O_2$ (µ = 0.11) in the as-deposited thin film closely resemble a binomial distribution, suggesting minor segregation. In contrast, the O (µ = 0.68) distribution deviates considerably, suggesting strong segregation, a seemingly contradictory outcome.

This difference in elemental distribution was further explored by analysing the variation in ionic composition of Fe, Ni, O, and $O_2$ in the as-deposited thin film at different LPEs as shown by **Figure 4a-e**. The Ga profile is also displayed to identify and isolate any effects from Ga ion beam damage near the surface. It is clear that the $O/O_2$ ratio decreases with increasing LPE, which is expected as the likelihood of molecular ion evaporation increases with LPE [69]. But it can also be seen that the $O/O_2$ ratio also changes with depth at the same LPE – increasing along the evaporation direction, while the concentration of the other ions stays uniform throughout the evaporation. This effect is particularly pronounced for O at higher LPEs (> 30 pJ), which explains the large deviation from a random distribution compared with the other ions. This is confirmed by the frequency distribution analysis of the as-deposited specimen measured at 20 pJ (**Figure 4f**), which denotes minor segregation for all ions, including O (µ = 0.28). Note that the homogeneous elemental mixing could also appear if a single-phase region was being probed. However, if that was the case, the measured Fe/Ni ratio would be different from the equiatomic value to reflect individual stoichiometry of the two oxide phases. Therefore, there is indeed substantial elemental mixing across the two phases in the as-deposited film, and the observed O ion considerable deviation from randomness in **Figure 3c** is an experimental artefact that depends on LPE.

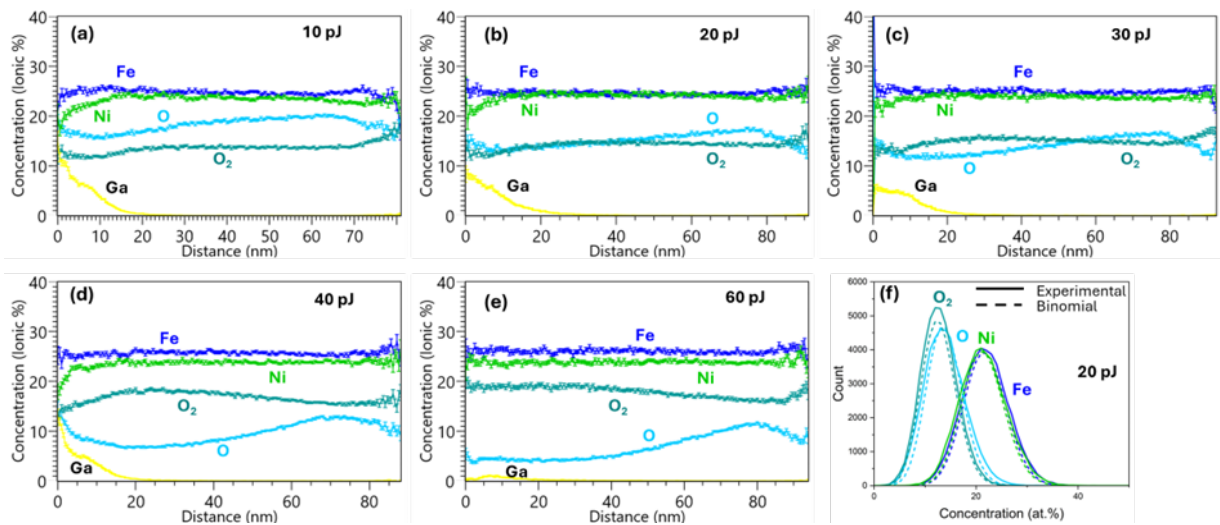

*Figure 4.* a-e) 1D ionic concentration profile along z-axis through the reconstruction of the as-deposited film analysed at different LPEs. f) Frequency distribution analysis of the specimen analysed at a LPE of 20 pJ. Here, the $O_2$ ions were displayed with a different colour for visual clarity.

The $O^+/(O^+ + O_2^+)$ ion ratio has been used as a direct measure of the surface field strength [70], and changes in this ratio signify changing field behaviour during the analysis. This change could have been triggered, at least partly, by local magnification effects as will be discussed in later sections. It is likely that at high DC voltage (increased with depth), the field near high curvature protrusions becomes high enough to allow evaporation of $O^+$ over molecular $O_2^+$ ions, or field-driven dissociation into $O^+$. The effect is less pronounced at low LPEs (high fields) because the likelihood of molecular ion evaporation is relatively lower [69].

As also denoted by XRD and SEM analyses in the previous sections, upon reduction at 280°C for 7 min, the oxide film starts to transform (**Figure 1a** and **Figure 2d-g**). Owing to the observed microstructural heterogeneity, different stages of the oxide transformation can be seen at the nanoscale, depending on the local reduction degree, as shown by APT analysis of specimens from the reduced (**Figure 2f**) regions of the film's microstructure (**Figure 5**, **Figure 6**, and **Figure 7**). Regions with a relatively low local reduction degree reveal that the beginning of the transformation is marked by Fe and Ni enrichment at interfacial regions (**Figure 5a-c** and **Figure 6c**), presumably prior oxide grain boundaries (GBs). 1D concentration profiles from numerous regions of interest (ROIs) across the interfaces (**Figure 5d**) show clear O deficiency at these interfacial regions (**Figure 5e-i**), with the most extensive reduction resulting in about 50% decrease in O content compared to the surrounding unreduced matrix/grain interior (**Figure 5g&h**). GBs were not visible in APT specimens of the as-deposited film analysed at various LPEs. This limited visibility is not solely due to crystallographic effects, such as field evaporation behaviour at GBs [71,72] which may influence preferential O loss. Instead, GBs became visible after reduction, as Fe and Ni segregate at the boundaries, forming Ni- and Fe-enriched interfacial regions compared to the grain interiors (**Figure 5b-c** and **Figure 6c**). This also suggests that reduction occurs first at the GBs.

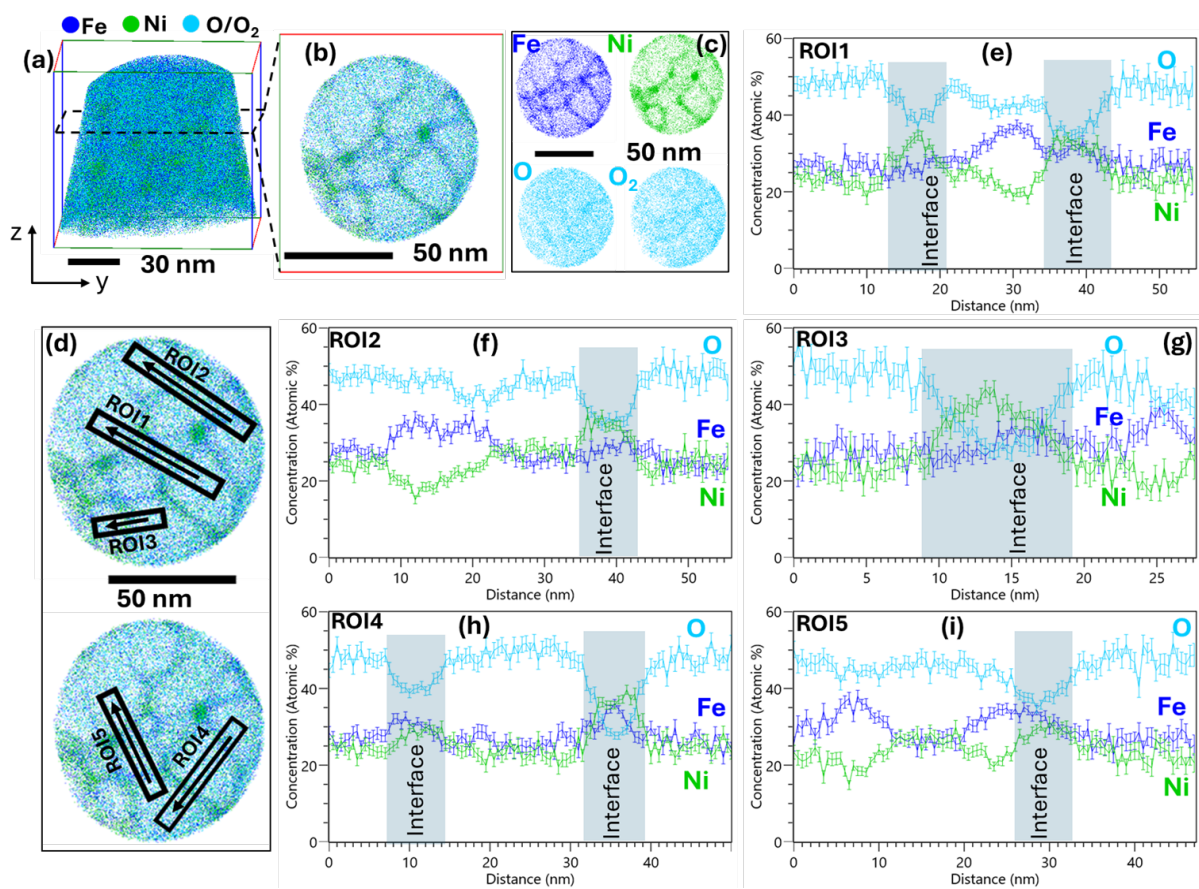

*Figure 5. APT analysis of the thin film after reduction at 280°C for 7 min: a) 3D reconstruction, b) a 5 nm slice of the reconstruction (top view) and c) its corresponding Fe, Ni, O, and $O_2$ ion maps. d) Different regions of interest (ROIs) from the slice used for 1D concentration profiles shown in e-i, suggesting that the reduction begins with O loss from prior oxide grain boundaries resulting in the formation of O-deficient interfacial regions (interfaces that are perpendicular to the ROIs are highlighted). ROI 1: 47.4×7×10 nm$^3$, ROI 2: 62.5×7×10 nm$^3$, ROI 3: 27.7×5×10 nm$^3$, ROI 4: 50.2×7×10 nm$^3$, ROI 5: 47.4×7×10 nm$^3$.*

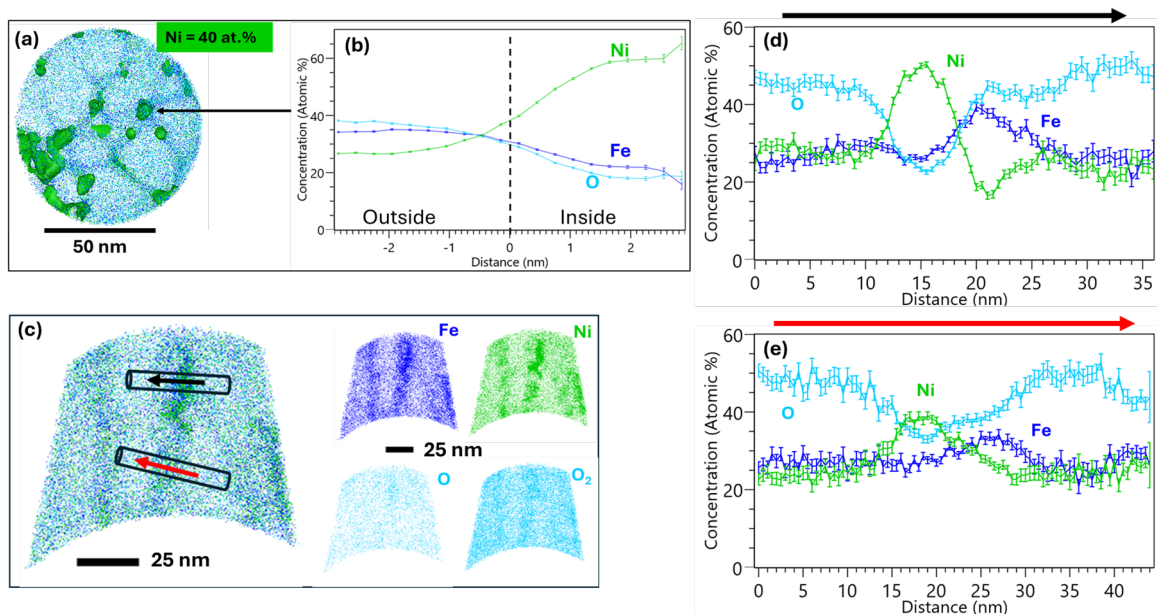

**Figure 6**. Ni enrichment at grain boundaries and triple junctions at the beginning of the reduction a) a 5 nm slice of the reconstruction of the film after 7 min reduction at 280°C, also shown in **Figure 5b**, with a 40 at.% Ni iso-surface. b) a proximity histogram from one of the triple junctions shown with the iso-surface. c) a 10 nm slice of the reconstruction shown in **Figure 5a** with its corresponding ion maps. (d) and (e) 1D compositional profiles measured from (d) 10×10×36.2 nm$^3$ and (e) 10×10×44 nm$^3$ cylindrical regions shown in (c).

The extent of Fe and Ni segregation to the GBs was further explored using both Ni (40 at.%) and Fe (40 at.%) isoconcentration surfaces (iso-surfaces) as shown by **Figure 6** and **Figure S5** (supplementary materials) respectively. The Ni iso-surface (**Figure 6a**) once again highlights substantial Ni enrichment at the GBs particularly triple junctions. A proximity histogram (proxigram) at one of the triple junctions (**Figure 6b**) exhibits a Ni concentration that is about 3 times higher than that of Fe. More importantly it shows that the Ni-rich triple junction is significantly O-deficient. On the other hand, proxigrams from several regions highlighted by the 40 at.% Fe iso-surface (**Figure S5** supplementary materials) show that the Fe-rich regions are also enriched with O but have a lower Ni concentration. This suggests that the initial reduction step marked by O-depletion from prior oxide GBs, results in the nucleation of the first metallic phase, which is largely Ni-rich, at the GBs and triple junctions, leaving behind an Fe-rich remaining oxide. Similarly, 1D concentration profiles across the GBs in a small slice along the long axis of the reconstruction (**Figure 6c-e**), shows that, in addition to the O deficiency, there is indeed substantial Ni segregation at the GBs (the highest being near the surface of the film) leaving a Ni-deficient, Fe and O-rich region nearby.

In the same manner, extended nanoscale transformation is seen from regions with a relatively higher reduction degree as shown by **Figure 7**. An O (25 at.%) iso-surface (**Figure 7b**) as well as ion maps (**Figure 7c&e**) from two different slices (**Figure 7b&d**) of the reconstruction (**Figure 7a**) reveal that the extensive reduction leads to nanoscale phase separation into two prominent parts: 1) regions that are enriched in Ni but substantially O-deficient, and 2) regions that are O-rich but highly Ni-deficient. Fe is dispersed within the two regions. This demonstrates that the metallic Ni-rich phase that nucleated at prior oxide GBs and triple junctions has grown into grain interiors, with a transformed remaining oxide ahead of the reaction front.

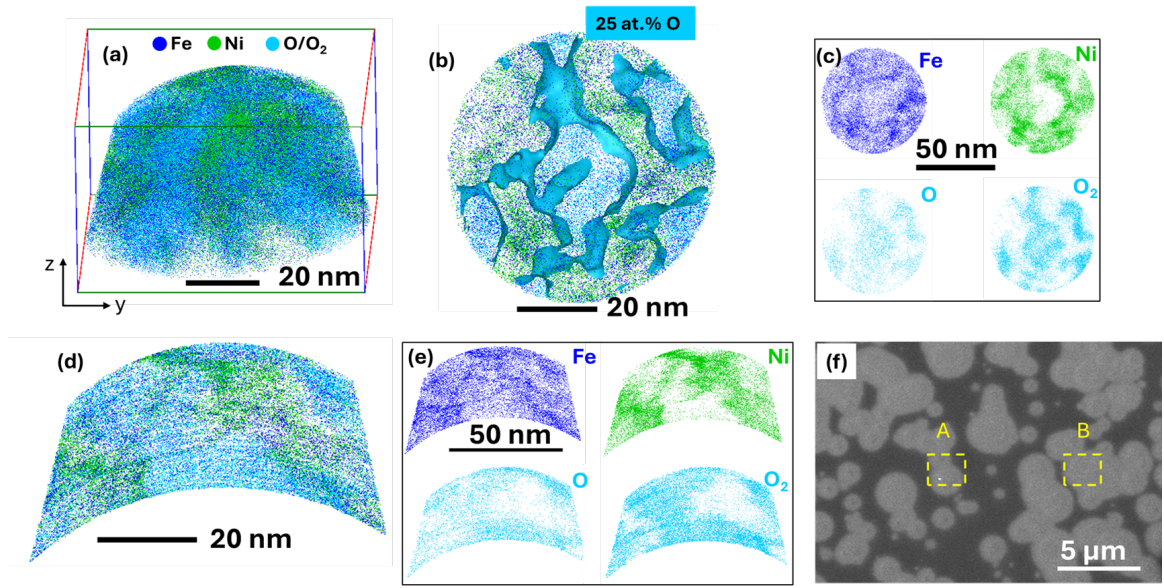

*Figure 7.* a-e) APT results of the thin film from a region with a higher reduction degree after reduction at 280°C for 7 min: a) 3D reconstruction. b) a 5 nm slice at the centre of z-axis of the reconstruction (top view), with a 25 at.% O isosurface; and d) a 10 nm slice at the centre of the x-axis of the reconstruction (side view) with their corresponding ion maps in c) and e) respectively, showing clear phase separation at the nanoscale in the initial stages of the reduction. f) Secondary electron SEM image showing the areas of interest (yellow dashed rectangles) where APT specimens in this figure (relatively higher reduction degree, B) and **Figure 5** (relatively lower reduction degree, A) were taken from.

Compositional profiles across various interfaces were used to assess the Fe partitioning in each of the two regions as shown in **Figure 8** and **Figure 9**. The figures show that, in general, the O-rich, Ni-deficient regions are even richer in Fe, and the Ni-rich, O-deficient regions have a relatively lower Fe content. Denoting again that the reduced metallic phase is Ni-rich, while the remaining oxide is Fe-rich. The Fe-rich remaining oxide phase shows a consistent concentration across the profiles (**Figure 8b-c** and **Figure 9b-c**), with Fe and O concentrations of approximately 50-60 at.% and 35-45 at.% respectively. Moreover, the Ni concentration in this phase is mostly < 10 at.% (**Figure 8c**, **Figure 9c**), substantially lower than the Ni concentration in the as-deposited state (~ 26 at.%, 50 pJ LPE). Implying that the remaining oxide is also significantly transformed being almost entirely Ni-free (Fe-O). On the other hand, the Ni-rich phase shows varying concentration profiles (**Figure 8** and **Figure 9**) marked by different Ni/Fe ratios, the highest being close to 3 (**Figure 9b**). This indicates the formation of several nanoscale transient phases, presumably related to their relative distance from the nucleation site and reaction front, as the reduction did not reach completion. Equivalently, the O concentration in the highly reduced instances of the Ni-rich phase is also very low (< 10 at.%, **Figure 9b-d**) compared to the as-deposited state (~ 47 at.%, 50 pJ LPE). This is consistent with the XRD observation that after extended reduction (10 min), NiO peaks were no longer present (**Figure 1a**), and signifies near-complete transformation to metal in some parts of the microstructure after just 7 min reduction. It is tempting to label the Ni-rich regions as just metallic Ni mixed with some remaining oxide; however, some patches of the compositional profiles labelled FeNi in **Figure 9b-d**, show that this is not the case, as the Fe concentration in these regions is as high as that of Ni whereas their O concentration is very low. Instead, alloying is simultaneously happening and the formation of nearly equiatomic FeNi in those regions suggests extensive Ni-Fe interdiffusion, an unusual occurrence at low temperatures such as the reduction temperature used.

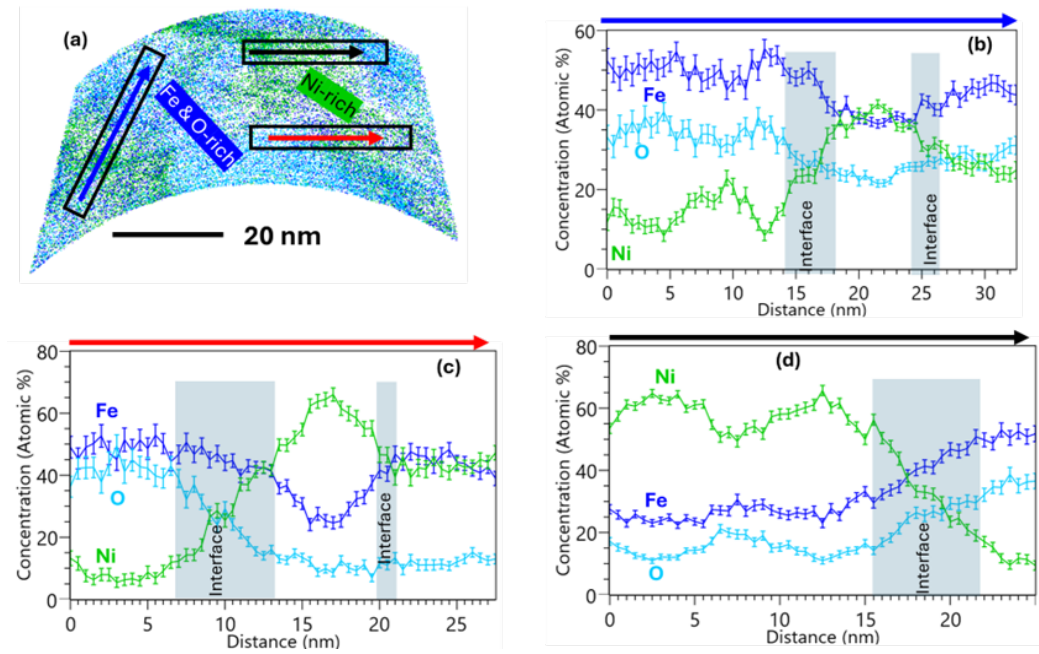

*Figure 8. (b-d) Composition profiles through (b) 10×4×32.8 nm$^3$, c) 10×4×27.9 nm$^3$ and d) 10×10×25.4 nm$^3$ regions of a 10 nm slice of the reconstruction (a) also shown in **Figure 7d**, showing that phase separation at the nanoscale in the initial stages of the reduction results into regions that are rich in Ni and O deficient, and others that are Fe rich with a relatively higher O concentration.*

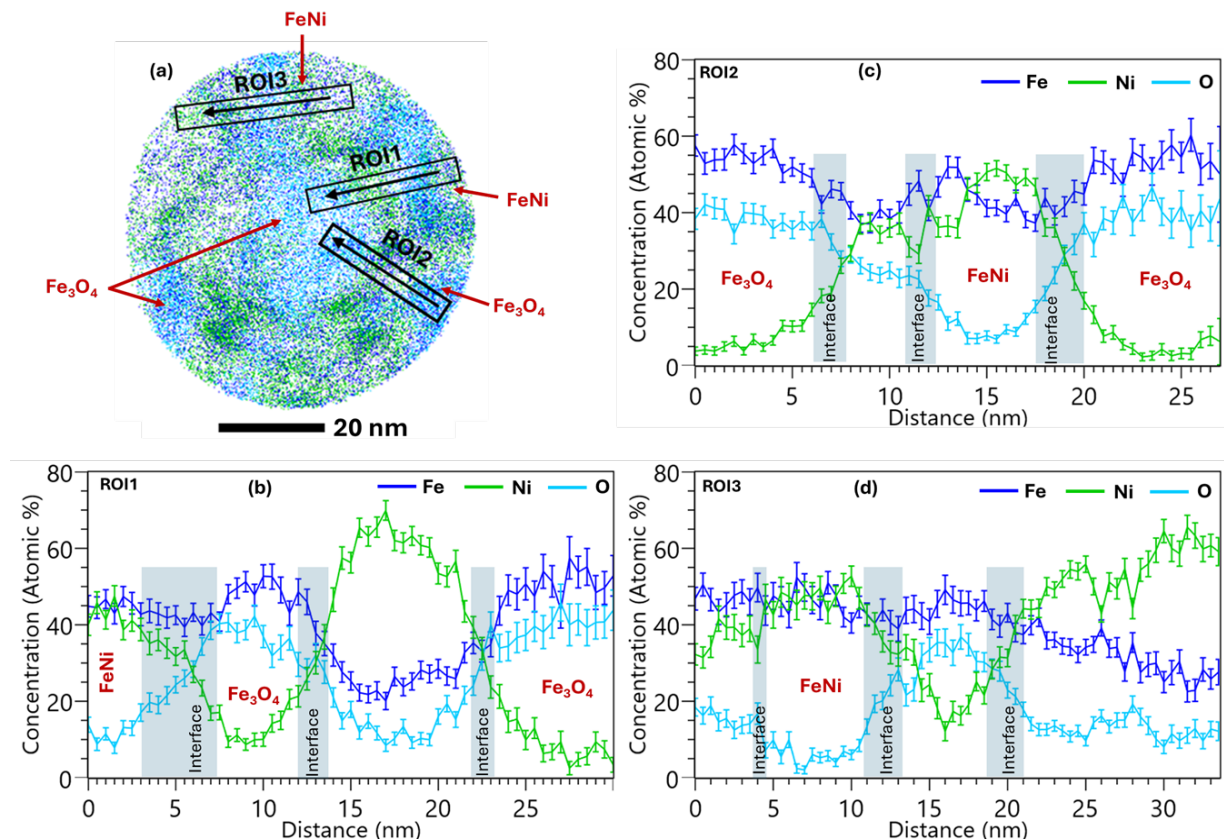

*Figure 9. (b-d) Composition profiles through different regions of interest (ROIs) in the 5 nm slice (a) also shown in **Figure 7b**, showing again that phase separation resulted in Ni-rich and O deficient regions as well as Fe rich regions with a relatively higher O concentration (Fe$_3$O$_4$). The profiles also display some zones (FeNi) with a metallic concentration of more than 90 at.%, suggesting potential Fe-Ni alloy formation. ROI 1: 5×30.4×5 nm$^3$, ROI 2: 5×27.1×5 nm$^3$, ROI 3: 5×34.6×5 nm$^3$, ROI 4: 5×28.8×5 nm$^3$, ROI 5: 5×33.8×5 nm$^3$.*

## 4  Discussion

### 4.1 Apparent nanoscale phase mixing in the as-deposited thin film

Based on XRD and Raman spectroscopy results, two types of oxides were identified, confirming that the as-deposited film is a two-phase mixture of $NiFe_2O_4$ and NiO. However, APT results revealed extensive nanoscale elemental mixing across the two phases of the as-deposited film into a nearly homogenous mixture. This is unexpected considering that Fe is only part of the spinel ($NiFe_2O_4$) phase and the stoichiometric Ni fraction in the spinel and NiO phases is 14.3 and 50 at.% respectively. Hence, if only single-phase (spinel) regions were probed, the resulting Fe/Ni concentration ratio would be around 2/1 not the observed 1/1.

Several effects could explain this discrepancy:

First, a central assumption in the reconstruction of the atom probe dataset is that the needle-shaped specimen consists of a perfectly hemispherical cap and that this geometry is maintained throughout the evaporation [71]. However, this is not always the case and often deviations from hemispherical shape happen around microstructural heterogeneities leading to uneven magnification and thus aberrations in the reconstruction [73], an issue known as local magnification effects (LMEs). This is particularly aggravated for multiphase materials with phases of largely dissimilar evaporation fields [74]. Due to these effects, a small region around the phase boundary is subjected to overlaps in trajectories of atoms from the different phases leading to inaccuracies in spatial atomic positions and elemental concentration.

The size of the trajectory overlapped region was estimated to be ~ 1 nm by Vurpillot et al. [75] in a simulation involving small (8 interatomic distances in size) spherical precipitates with an evaporation field of 0.85 – 1.15 times that of the matrix. A value of 1.4 – 2 nm was also measured experimentally [76] for 7 – 12 nm $Al_3Zr$ precipitates in a 7050 Al alloy. Generally, significant compositional inaccuracies are expected for particles < 5 nm in size [75,77]. Using the non-overlapping spinel peak at 35.45° 2θ (**Figure 1a**), and assuming minimal residual strain contribution to peak broadening, the crystallite size in the as-deposited film (**Figure S1** supplementary materials) was estimated using the Scherrer equation to be ~11 nm. Generally, the crystallite size estimated by XRD provides a lower bound for the grain size. The grain size of around 20 nm, as revealed by APT after 7 minutes of reduction (**Figure 5e&h**), likely represents an upper bound, considering slight grain growth during reduction at 280°C. Even though grain size and morphology are strongly influenced by reconstruction parameters, such as initial tip diameter and shank angle for fixed shank angle reconstructions, the resulting grain size and morphology (**Figure 5e&h**) are consistent with those observed in other thin film alloy systems, typically exhibiting 10~20 nm equiaxed grains perpendicular to the film growth direction and elongated grains along the growth direction [78]. Thus, the crystallite size of the two oxides should be between 11–20 nm, which is much larger than the effective zone of the local magnification effect, further suggesting that the observed phase mixing in APT cannot be fully attributed to LME.

Indeed, laser-assisted field evaporation of non-metallic materials such as oxides is a complex process [71], and as mentioned, involves the evaporation and dissociation of molecular ions, and a large fraction of multiple events. The formation of molecular ions requires reconfiguration of surface atoms [79] which also disrupt the hemispherical geometry and enhance local magnification effects. The time needed for atomic reconfiguration could also result in delayed evaporation of some species [53]. A portion of the multiple events could also originate from correlated evaporation where multiple adjacent atoms evaporate simultaneously due to enhanced evaporation probability after one atom is desorbed [80]. This would also contribute to the surface field inhomogeneity. The resulting field gradients could then drive surface migration/diffusion [81,82] of some of the atoms degrading further the spatial resolution. Even though Oberdorfer et al. [83] found this thermally-activated process to be minimal in NiO, it could be important here considering the presence of a second oxide phase and a highly varying surface curvature. Moreover, as also previously mentioned, multiple events that are closer in space than the detector dead zone or in time than the detector dead time (~3 ns for the LEAP 5000 XR [84]) will not only affect the composition but also decrease the reconstruction accuracy due to pile-up effects [61,84]. Overall, the synergy between some or all these factors is believed to be the source of the observed phase mixing. The exact reasons were not explored further in this study.

## 4.2 Incubation time and grain boundary nucleation

The results also revealed that the reduction of the oxide thin film at 280°C does not start before an incubation time of more than 5 min, and that after this time, reduction starts at prior oxide grain boundaries and triple junctions (**Figure 5** and **Figure 6**). This behaviour is characteristic for a nucleation-limited transformation [85,86], underscoring the critical role of heterogeneous nucleation at these reduction conditions.

Generally, reduction of bulk or nanocrystalline Ni, Fe, or Fe-Ni oxides is sluggish at temperatures below about 250-350°C [86–90]. The employed reduction temperature of 280°C represents a low-to-moderate activation regime (**Figure S7** in supplementary materials), which accounts for the observed incubation time and incomplete conversion. Manukyan et al. [91] made similar observations in their study on the reduction of bulk NiO (grain size of 1-3 μm) using $H_2$ at temperatures ranging from 270 to 1320°C. They found that at 270-500°C, the reaction kinetics followed the Avrami nucleation model with an Avrami exponent of <1 (0.58 at 335°C). The transformation was preceded by long induction times (2000 s at 270°C), and full conversion was not attained. They also found that at these relatively low temperatures, nucleation started at the surface and NiO grain boundaries [91].

The formation of a denser metallic phase upon reduction induces significant volumetric shrinkage (e.g.: lattice parameters of Ni/NiO differ by 17% [92]) resulting in large strains in the surrounding oxide matrix. This favours heterogeneous nucleation and subsequent growth at free surfaces such as pre-existing pores, prior oxide grain and phase boundaries, and other defects as they facilitate elastic strain energy relaxation, thereby lowering the nucleation barrier [21,22,91]. These defects also provide easier paths for diffusion which is crucial at low temperatures [22]. It has also been reported [86] that the presence of O vacancies promotes efficient $H_2$ adsorption-dissociation on Ni atoms during hydrogen reduction of NiO, and that the induction time is associated with the formation of sufficient defects (O vacancies). It is likely that a higher equilibrium concentration of O vacancies existed in the GB regions of the as-deposited thin film oxide than the grain interiors. Consequently, the transformation starts from thereon (**Figure 5**&**Figure 6**) and variations/heterogeneity in reduction degree across the microstructure depend on the local defect concentration (**Figure S2**).

### 4.3 Nanoscale phase separation and low temperature alloying

The reduction sequence of a multicomponent oxide follows similar steps as those of its constituent phases. For Fe-Ni-O, this is dictated by the reduction sequence of both NiO and $NiFe_2O_4$. NiO is a model oxide [91] whose reduction only occurs in a single step (NiO → Ni), whereas that of $NiFe_2O_4$ takes several steps ($NiFe_2O_4$ → NiO + $Fe_3O_4$ → Ni + $Fe_3O_4$ → $Fe_xNi_y$ + FeO → Ni-Fe alloy) [27,93] depending on the temperature and hydrogen partial pressure. APT results demonstrated that the GB nucleation (**Figure 6**) and the ensuing growth (**Figure 8** and **Figure 9**) were accompanied by significant nanoscale elemental partitioning, forming an O-depleted (< 10 at.% in highly reduced regions) Ni-rich phase, and leaving behind an Fe-O rich phase that is nearly devoid of Ni (<10 at.%) in the vicinity of the metal/oxide interface. Notably, the impact of molecular dissociation and multiple events were substantially reduced in this case, as multiple events were only 14.1% for the highly reduced specimen shown in **Figure 7a**, down from over 40% in the as-deposited state.

The O-deficient, Ni-rich phase with varying Ni/Fe ratios (**Figure 8** and **Figure 9**) corresponds to the metallic FCC $Fe_xNi_y$ phase, as identified by XRD. The remaining but transformed Fe-O rich phase represents the Fe spinel, $Fe_3O_4$, as denoted by its consistent concentration in various regions of the specimen (**Figure 9b&c**). This is also in line with the XRD results as only spinel peaks remained after extensive 10 min reduction (**Figure 1a**). However, since there were other regions of the microstructure that remained completely unreduced/non-transformed after 7 min (**Figure 2f**), and both $Fe_3O_4$ and $NiFe_2O_4$ share the same spinel structure, no discernible shifts in spinel peaks were observed on the diffractogram. Besides, FeO is thermodynamically stable only above 570°C [94], well above the reduction temperature used. $Fe_3O_4$ is gradually being reduced into Fe which is simultaneously incorporated into the metallic phase forming alloys with different Ni/Fe ratios across different areas of the film depending on their local reduction degree. This sequence (NiO + $NiFe_2O_4$ → $Fe_xNi_y$ + $Fe_3O_4$ → FeNi) mirrors parts of the aforementioned individual oxide reduction sequences and aligns with the thermodynamic stability of NiO relative to Fe oxides in accordance with the Ellingham-Richardson diagram [1,6]. The relatively low (< 10 at.% in highly reduced regions) concentration of O in the metallic $Fe_xNi_y$ phase and the corresponding low (< 10 at.%) concentration of Ni in the $Fe_3O_4$ phase regions, suggests that the metallic phase is embedded with a few domains of the remaining oxide [91], and vice versa, reflecting the extent of local conversion. Metal interdiffusion near the metal-oxide interface is also possible [4].

Despite the relatively low reduction temperature, nanoscale results also show considerable Ni-Fe alloying in some regions of the reduced film (labelled FeNi in **Figure 9**). Lattice diffusion in the Fe-Ni system is extremely slow at low temperatures (below about 400°C [34]), which has hindered the exploration of the low temperature part of the Fe-Ni phase diagram [95]. Thin films [96] and nanostructuring [97,98] have been tried to bypass this limitation as nanostructures provide shorter distances for diffusion and have an increased proportion of grain boundaries which enhances the diffusion rate by several orders of magnitude compared to coarse-grained structures [99,100]. The effective interdiffusion coefficient of Fe and Ni in nanosized Fe/Ni multilayer thin films in the temperature range of 300 to 430°C, was determined by Liu and Barmak [101] to be in the range of $2.1 \times 10^{-19}$–$1.1 \times 10^{-17}$ cm$^2$/s. On the contrary, the diffusivity of Fe and Ni in the FeNi lattice, at temperatures slightly below 320°C, was estimated to be more than seven orders of magnitude lower, ranging between $10^{-26}$ to $10^{-27}$ cm$^2$/s [101,102]. Using both values of the diffusion coefficients (D), the diffusion distance $x=\sqrt{6Dt}$ [103] after 7 min reduction was estimated to be 0.19 - 1.36 nm for the nanosized Fe/Ni interdiffusion and $1.3 \times 10^{-5}$ – $4.1 \times 10^{-5}$ nm for the lattice diffusion. The size of the observed FeNi alloy regions (labelled FeNi in **Figure 9**) is about 5 -10 nm, which is comparable to the estimated diffusion distance for nanosized Fe/Ni. Therefore, both the nanostructure and the nanometric thickness of the oxide film promoted alloying through enhanced Ni-Fe interdiffusion.

The addition of elements with a high propensity for alloying (near zero mixing enthalpy) and whose oxides are thermodynamically less stable such as Ni, has also been found to enhance the reduction kinetics of Fe oxides [6,87]. Once Ni is reduced, subsequent alloying lowers the activity of Fe near the metal-oxide interface [4] thereby shifting the Fe-O-$H_2$ equilibrium towards metallic Fe. For instance, Pan et al. [6] studied the influence of secondary elements, including Ni, on the reduction of Fe oxides nanoparticles with temperatures of up to 900°C, and demonstrated that the presence of Ni (as $NiFe_2O_4$) shortens the reduction pathway of $Fe_3O_4$ into a one-step process where the oxide transforms directly into alloy ($NiFe_2O_4 \rightarrow Fe_3O_4$ + FeNi $\rightarrow$ FeNi), bypassing the FeO stage when compared to the reduction pathway of $Fe_3O_4$ at the same conditions. Furthermore, the onset and final transformation temperatures were lower compared to both NiO and $Fe_3O_4$ [6]. This was attributed to alloying and autocatalytic effects induced by the Ni through hydrogen spillover. Similarly, Zhang et al. [27] studied the reduction of several bulk Fe-Ni-O compounds and pointed out that the hydrogen reduction of $NiFe_2O_4$ reached completion at 513°C, and hence FeO was also not observed. Ultimately, the extensive reduction and low temperature alloying stems from the combined influence of the film's nanostructure, and Ni autocatalytic effects.

## 5 Conclusions

Early-stages of the hydrogen direct reduction of Fe-Ni-O thin films was investigated using an equiatomic Fe-Ni oxide thin film consisting of two phases, $NiFe_2O_4$ + NiO. The oxide film was isothermally held in a flow of pure $H_2$ at 280°C for 5, 7, or 10 min, and near-atomic scale transformations were monitored using APT. It is revealed that:

1. The transformation was nucleation-limited with the first metallic phase forming at prior oxide grain boundaries preceded by an incubation time of more than 5 min.
2. Nucleation and subsequent growth results in phase separation into a Ni-rich $Fe_xNi_y$ alloy and a Fe-rich oxide ($Fe_3O_4$). Extended reduction induces direct $Fe_3O_4$ reduction and alloying into nearly equiatomic FeNi.
3. The extensive low-temperature reduction and alloying was facilitated by the film's nanostructure and Ni autocatalytic effects through alloying and hydrogen spillover.

**Acknowledgments**

The authors are grateful to W. Xia, L. Alfes, and S. Delsing for assistance with reduction experiments. E.U thanks Shiv Shankar for some fruitful discussions. E.U. acknowledges funding from the International Max Planck Research School for Sustainable Metallurgy (IMPRS SusMet). S.L.W. acknowledges financial support from the Alexander von Humboldt Fellowship (hosted by D.R.). S.L.W. and D.R. also thank the MPG-FhG Cooperation Grant MaRS. The Center for Interface-Dominated High Performance Materials (Zentrum für Grenzflächendominierte Höchstleistungswerkstoffe, ZGH) is acknowledged for the use of its facilities. The Laboratory of Industrial Chemistry, Ruhr University Bochum is acknowledged for using its furnace lab for reduction experiments. The Chair of Applied Laser Technologies (LAT), Ruhr University Bochum is acknowledged for the use of Raman microscope.

**References**

[1] S. Wei, Y. Ma, D. Raabe, One step from oxides to sustainable bulk alloys, Nature 633 (2024) 816–822. https://doi.org/10.1038/s41586-024-07932-w.

[2] S. Shankar, B. Ratzker, Y. Ma, D. Raabe, Hydrogen-based direct reduction of multicomponent oxides: Insights from powder and pre-sintered precursors toward sustainable alloy design, Acta Mater. 301 (2025) 121528. https://doi.org/10.1016/j.actamat.2025.121528.

[3] S. Shankar, B. Ratzker, C. Pistidda, D. Raabe, Y. Ma, Unveiling Hydrogen-Based Direct Reduction Mechanisms of Multicomponent Oxides via In Situ High-

Energy X-ray Diffraction, ACS Sustain. Chem. Eng. 14 (2026) 1762–1768. https://doi.org/10.1021/acssuschemeng.5c12301.

[4] S. Shankar, B. Ratzker, A.K. da Silva, T.M. Schwarz, H. Brouwer, B. Gault, Y. Ma, D. Raabe, Unraveling the thermodynamics and mechanism behind the lowering of direct reduction temperatures in oxide mixtures, Mater. Today 90 (2025) 43–51. https://doi.org/10.1016/j.mattod.2025.08.021.

[5] C. Kenel, T. Davenport, X. Li, R.N. Shah, D.C. Dunand, Kinetics of alloy formation and densification in Fe-Ni-Mo microfilaments extruded from oxide- or metal-powder inks, Acta Mater. 193 (2020) 51–60. https://doi.org/10.1016/j.actamat.2020.04.038.

[6] Y. Pan, X. Liu, S. Zhen, J. Zhao, Y. Wang, M. Ge, J. Zhang, Y. Pan, L. Gu, L. Zhang, D. Zhou, D. Su, Alloying Effects on Iron Oxide Redox Pathways: Insights into Sustainable Hydrogen-Based Reduction, J. Phys. Chem. Lett. 16 (2025) 5506–5514. https://doi.org/10.1021/acs.jpclett.5c00954.

[7] J. Lee, K.-H. Huang, K.-C. Hsu, H.-C. Tung, J.-W. Lee, J.-G. Duh, Applying composition control to improve the mechanical and thermal properties of Zr–Cu–Ni–Al thin film metallic glass by magnetron DC sputtering, Surf. Coat. Technol. 278 (2015) 132–137. https://doi.org/10.1016/j.surfcoat.2015.07.015.

[8] S.-S. Lin, C.-S. Liao, Structure and physical properties of W-doped $HfO_2$ thin films deposited by simultaneous RF and DC magnetron sputtering, Surf. Coat. Technol. 232 (2013) 46–52. https://doi.org/10.1016/j.surfcoat.2013.04.051.

[9] A. Furlan, D. Grochla, Q. D'Acremont, G. Pernot, S. Dilhaire, A. Ludwig, Influence of Substrate Temperature and Film Thickness on Thermal, Electrical, and Structural Properties of HPPMS and DC Magnetron Sputtered Ge Thin Films, Adv. Eng. Mater. 19 (2017) 1600854. https://doi.org/10.1002/adem.201600854.

[10] A.A. Turkin, Y.T. Pei, K.P. Shaha, C.Q. Chen, D.I. Vainshtein, J.Th.M. De Hosson, On the evolution of film roughness during magnetron sputtering deposition, J. Appl. Phys. 108 (2010) 094330. https://doi.org/10.1063/1.3506681.

[11] E.L. Miller, D. Paluselli, B. Marsen, R.E. Rocheleau, Low-temperature reactively sputtered iron oxide for thin film devices, Thin Solid Films 466 (2004) 307–313. https://doi.org/10.1016/j.tsf.2004.02.093.

[12] J. Jia, H. Yamamoto, T. Okajima, Y. Shigesato, On the Crystal Structural Control of Sputtered $TiO_2$ Thin Films, Nanoscale Res. Lett. 11 (2016) 324. https://doi.org/10.1186/s11671-016-1531-5.

[13] L. Patterer, E.B. Mayer, S. Mráz, P.J. Pöllmann, M. Hans, D. Primetzhofer, I.R. Souza Filho, H.J. Springer, J.M. Schneider, Effect of Si on the hydrogen-based direct

reduction of Fe2O3 studied by XPS of sputter-deposited thin-film model systems, Scr. Mater. 233 (2023) 115515. https://doi.org/10.1016/j.scriptamat.2023.115515.

[14] F. Lourens, E. Suhr, A. Schnickmann, T. Schirmer, A. Ludwig, High-Throughput Study of the Phase Constitution of the Thin Film System Mg–Mn–Al–O, Adv. Eng. Mater. 26 (2024) 2302091. https://doi.org/10.1002/adem.202302091.

[15] T.H. Piotrowiak, O.A. Krysiak, E. Suhr, J. Zhang, R. Zehl, A. Kostka, W. Schuhmann, A. Ludwig, Sputter-Deposited La–Co–Mn–O Nanocolumns as Stable Electrocatalyst for the Oxygen Evolution Reaction, Small Struct. 5 (2024) 2300415. https://doi.org/10.1002/sstr.202300415.

[16] A. Mockute, A. Ludwig, High-throughput exploration of the Na2O-ZrO2-SiO2-P2O5 composition space using thin film material libraries, J. Mater. Sci. 59 (2024) 16900–16908. https://doi.org/10.1007/s10853-024-10221-6.

[17] E.J.J. Martin, M. Yan, M. Lane, J. Ireland, C.R. Kannewurf, R.P.H. Chang, Properties of multilayer transparent conducting oxide films, Thin Solid Films 461 (2004) 309–315. https://doi.org/10.1016/j.tsf.2004.01.103.

[18] S.A. Chambers, Epitaxial Growth and Properties of Doped Transition Metal and Complex Oxide Films, Adv. Mater. 22 (2010) 219–248. https://doi.org/10.1002/adma.200901867.

[19] F. Lourens, D. Rogalla, E. Suhr, A. Ludwig, The Influence of Annealing on the Compositional and Crystallographic Properties of Sputtered Li–Al–O Thin Films, Adv. Eng. Mater. 26 (2024) 2400602. https://doi.org/10.1002/adem.202400602.

[20] A. Sarkar, T.L. Schanche, M. Wallin, J. Safarian, Evaluating the Reaction Kinetics on the H2 Reduction of a Manganese Ore at Elevated Temperatures, J. Sustain. Metall. 10 (2024) 2085–2103. https://doi.org/10.1007/s40831-024-00964-6.

[21] Y. Ma, I.R. Souza Filho, X. Zhang, S. Nandy, P. Barriobero-Vila, G. Requena, D. Vogel, M. Rohwerder, D. Ponge, H. Springer, D. Raabe, Hydrogen-based direct reduction of iron oxide at 700°C: Heterogeneity at pellet and microstructure scales, Int. J. Miner. Metall. Mater. 29 (2022) 1901–1907. https://doi.org/10.1007/s12613-022-2440-5.

[22] S.-H. Kim, X. Zhang, Y. Ma, I.R. Souza Filho, K. Schweinar, K. Angenendt, D. Vogel, L.T. Stephenson, A.A. El-Zoka, J.R. Mianroodi, M. Rohwerder, B. Gault, D. Raabe, Influence of microstructure and atomic-scale chemistry on the direct reduction of iron ore with hydrogen at 700°C, Acta Mater. 212 (2021) 116933. https://doi.org/10.1016/j.actamat.2021.116933.

[23] A.P. Arun, N. Sreenivasan, J.H. Patil, R. Kusanur, H.L. Ramachandraiah, M. Ramakrishna, Thin Films for Next Generation Technologies: A Comprehensive Review

of Fundamentals, Growth, Deposition Strategies, Applications, and Emerging Frontiers, Processes 13 (2025) 3846. https://doi.org/10.3390/pr13123846.

[24] A.L. Greer, Diffusion and reactions in thin films, Appl. Surf. Sci. 86 (1995) 329–337. https://doi.org/10.1016/0169-4332(94)00399-8.

[25] M. Häger, S. Keilholz, H. Kohlmann, The Reduction of Group 6–8 Transition Metal Oxides with Hydrogen—From Ore Smelting to Reaction Pathways#, Eur. J. Inorg. Chem. 28 (2025) e202500089. https://doi.org/10.1002/ejic.202500089.

[26] W. Jin, P. Sharma, P. Singh, A. Kundu, G. Balasubramanian, H.M. Chan, Solid State Reduction Driven Synthesis of Mn Containing Multi-principal Component Alloys, Metall. Mater. Trans. A 55 (2024) 3799–3808. https://doi.org/10.1007/s11661-024-07490-w.

[27] Y. Zhang, W. Wei, X. Yang, F. Wei, Reduction of Fe and Ni in Fe-Ni-O systems, J. Min. Metall. Sect. B Metall. 49 (2013) 13–20. https://doi.org/10.2298/JMMB120208038Z.

[28] C. Yang, L. Wei, D. Zhu, J. Pan, G. Xia, S. Qu, Hydrogen-based direct reduction behavior of iron ore pellets with iron grades ranging from 59 % to 68 %, Int. J. Hydrog. Energy 138 (2025) 787–801. https://doi.org/10.1016/j.ijhydene.2025.04.006.

[29] J.A. Van Orman, K.L. Crispin, Diffusion in Oxides, Rev. Mineral. Geochem. 72 (2010) 757–825. https://doi.org/10.2138/rmg.2010.72.17.

[30] P. Kofstad, Defects and transport properties of metal oxides, Oxid. Met. 44 (1995) 3–27. https://doi.org/10.1007/BF01046721.

[31] A. Goswami, Thin Film Fundamentals, New Age International, 1996.

[32] T.H. Piotrowiak, R. Zehl, E. Suhr, L. Banko, B. Kohnen, D. Rogalla, A. Ludwig, Unusual Phase Formation in Reactively Sputter-Deposited La—Co—O Thin-Film Libraries, Adv. Eng. Mater. 25 (2023) 2201050. https://doi.org/10.1002/adem.202201050.

[33] P. Marx, S. Shukla, A.E.P. Mendoza, F. Lourens, C. Andronescu, A. Ludwig, Structural and electrocatalytic properties of La-Co-Ni oxide thin films, Thin Solid Films 836 (2026) 140880. https://doi.org/10.1016/j.tsf.2026.140880.

[34] C.-W. Yang, D.B. Williams, J.I. Goldstein, A revision of the Fe-Ni phase diagram at low temperatures (<400 °C), J. Phase Equilibria 17 (1996) 522–531. https://doi.org/10.1007/BF02665999.

[35] K. Thompson, D. Lawrence, D.J. Larson, J.D. Olson, T.F. Kelly, B. Gorman, In situ site-specific specimen preparation for atom probe tomography, Ultramicroscopy 107 (2007) 131–139. https://doi.org/10.1016/j.ultramic.2006.06.008.

[36]   S.B. Narang, K. Pubby, Nickel Spinel Ferrites: A review, J. Magn. Magn. Mater. 519 (2021) 167163. https://doi.org/10.1016/j.jmmm.2020.167163.

[37]   D. Levy, R. Giustetto, A. Hoser, Structure of magnetite (Fe3O4) above the Curie temperature: a cation ordering study, Phys. Chem. Miner. 39 (2012) 169–176. https://doi.org/10.1007/s00269-011-0472-x.

[38]   O.A. Restrepo, Ó. Arnache, N. Mousseau, An approach to understanding the formation mechanism of NiFe2O4 inverse spinel, Materialia 33 (2024) 102031. https://doi.org/10.1016/j.mtla.2024.102031.

[39]   F. Bosi, U. Hålenius, H. Skogby, Crystal chemistry of the magnetite-ulvöspinel series, Am. Mineral. 94 (2009) 181–189. https://doi.org/10.2138/am.2009.3002.

[40]   K.-N. Tu, ed., Elastic stress and strain in thin films, in: Electron. Thin-Film Reliab., Cambridge University Press, Cambridge, 2010: pp. 118–140. https://doi.org/10.1017/CBO9780511777691.007.

[41]   M. Huff, Review Paper: Residual Stresses in Deposited Thin-Film Material Layers for Micro- and Nano-Systems Manufacturing, Micromachines 13 (2022) 2084. https://doi.org/10.3390/mi13122084.

[42]   C. Guillén, J. Herrero, Influence of Acceptor Defects on the Structural, Optical and Electrical Properties of Sputtered NiO Thin Films, Phys. Status Solidi A 218 (2021) 2100237. https://doi.org/10.1002/pssa.202100237.

[43]   A. Ahlawat, V.G. Sathe, Raman study of NiFe2O4 nanoparticles, bulk and films: effect of laser power, J. Raman Spectrosc. 42 (2011) 1087–1094. https://doi.org/10.1002/jrs.2791.

[44]   J.-L. Ortiz-Quiñonez, U. Pal, M.S. Villanueva, Structural, Magnetic, and Catalytic Evaluation of Spinel Co, Ni, and Co–Ni Ferrite Nanoparticles Fabricated by Low-Temperature Solution Combustion Process, ACS Omega 3 (2018) 14986–15001. https://doi.org/10.1021/acsomega.8b02229.

[45]   V.G. Ivanov, M.V. Abrashev, M.N. Iliev, M.M. Gospodinov, J. Meen, M.I. Aroyo, Short-range B-site ordering in the inverse spinel ferrite NiFe2O4, Phys. Rev. B 82 (2010) 024104. https://doi.org/10.1103/PhysRevB.82.024104.

[46]   Z.H. Zhou, J.M. Xue, J. Wang, H.S.O. Chan, T. Yu, Z.X. Shen, NiFe2O4 nanoparticles formed in situ in silica matrix by mechanical activation, J. Appl. Phys. 91 (2002) 6015–6020. https://doi.org/10.1063/1.1462853.

[47]   X. Du, T.-L. Yao, Q. Wei, H. Zhang, Y. Huang, Investigation of Fe−Ni Mixed-Oxide Catalysts for the Reduction of NO by CO: Physicochemical Properties and Catalytic Performance, Chem. – Asian J. 14 (2019) 2966–2978. https://doi.org/10.1002/asia.201900782.

[48]    R.E. Dietz, G.I. Parisot, A.E. Meixner, Infrared Absorption and Raman Scattering by Two-Magnon Processes in NiO, Phys. Rev. B 4 (1971) 2302–2310. https://doi.org/10.1103/PhysRevB.4.2302.

[49]    A. Mendoza-Galván, M.A. Vidales-Hurtado, A.M. López-Beltrán, Comparison of the optical and structural properties of nickel oxide-based thin films obtained by chemical bath and sputtering, Thin Solid Films 517 (2009) 3115–3120. https://doi.org/10.1016/j.tsf.2008.11.094.

[50]    S. Cheemadan, M.C. Santhosh Kumar, Effect of substrate temperature and oxygen partial pressure on RF sputtered NiO thin films, Mater. Res. Express 5 (2018) 046401. https://doi.org/10.1088/2053-1591/aab875.

[51]    M. Mishra, S. Barthwal, B.R. Tak, R. Singh, Temperature-Driven Perturbations in Growth Kinetics, Structural and Optical Properties of NiO Thin Films, Phys. Status Solidi A 218 (2021) 2100241. https://doi.org/10.1002/pssa.202100241.

[52]    X. Zhou, Effect of Pore Formation on Redox-Driven Phase Transformation, Phys. Rev. Lett. 130 (2023). https://doi.org/10.1103/PhysRevLett.130.168001.

[53]    S.-H. Kim, S. Bhatt, D.K. Schreiber, J. Neugebauer, C. Freysoldt, B. Gault, S. Katnagallu, Understanding atom probe's analytical performance for iron oxides using correlation histograms and ab initio calculations, New J. Phys. 26 (2024) 033021. https://doi.org/10.1088/1367-2630/ad309e.

[54]    A. Devaraj, R. Colby, W.P. Hess, D.E. Perea, S. Thevuthasan, Role of Photoexcitation and Field Ionization in the Measurement of Accurate Oxide Stoichiometry by Laser-Assisted Atom Probe Tomography, J. Phys. Chem. Lett. 4 (2013) 993–998. https://doi.org/10.1021/jz400015h.

[55]    D. Santhanagopalan, D.K. Schreiber, D.E. Perea, R.L. Martens, Y. Janssen, P. Khalifah, Y.S. Meng, Effects of laser energy and wavelength on the analysis of LiFePO4 using laser assisted atom probe tomography, Ultramicroscopy 148 (2015) 57–66. https://doi.org/10.1016/j.ultramic.2014.09.004.

[56]    H.S. Stein, S. Zhang, Y. Li, C. Scheu, A. Ludwig, Photocurrent Recombination Through Surface Segregation in Al–Cr–Fe–O Photocathodes, Z. Für Phys. Chem. 234 (2020) 605–614. https://doi.org/10.1515/zpch-2019-1459.

[57]    Y. Li, D. Zanders, M. Meischein, A. Devi, A. Ludwig, Investigation of an atomic-layer-deposited Al2O3 diffusion barrier between Pt and Si for the use in atomic scale atom probe tomography studies on a combinatorial processing platform, Surf. Interface Anal. 53 (2021) 727–733. https://doi.org/10.1002/sia.6955.

[58]    H. Waldl, M. Hans, M. Schiester, D. Primetzhofer, M. Burtscher, N. Schalk, M. Tkadletz, Decomposition of CrN induced by laser-assisted atom probe tomography, Ultramicroscopy 246 (2023) 113673. https://doi.org/10.1016/j.ultramic.2022.113673.

[59]    L. Mancini, N. Amirifar, D. Shinde, I. Blum, M. Gilbert, A. Vella, F. Vurpillot, W. Lefebvre, R. Lardé, E. Talbot, P. Pareige, X. Portier, A. Ziani, C. Davesnne, C. Durand, J. Eymery, R. Butté, J.-F. Carlin, N. Grandjean, L. Rigutti, Composition of Wide Bandgap Semiconductor Materials and Nanostructures Measured by Atom Probe Tomography and Its Dependence on the Surface Electric Field, J. Phys. Chem. C 118 (2014) 24136–24151. https://doi.org/10.1021/jp5071264.

[60]    F. Tang, B. Gault, S.P. Ringer, J.M. Cairney, Optimization of pulsed laser atom probe (PLAP) for the analysis of nanocomposite Ti–Si–N films, Ultramicroscopy 110 (2010) 836–843. https://doi.org/10.1016/j.ultramic.2010.03.003.

[61]    S. Pedrazzini, A.J. London, B. Gault, D. Saxey, S. Speller, C.R.M. Grovenor, M. Danaie, M.P. Moody, P.D. Edmondson, P.A.J. Bagot, Nanoscale Stoichiometric Analysis of a High-Temperature Superconductor by Atom Probe Tomography, Microsc. Microanal. 23 (2017) 414–424. https://doi.org/10.1017/S1431927616012757.

[62]    O. Cojocaru-Mirédin, Y. Yu, J. Köttgen, T. Ghosh, C.-F. Schön, S. Han, C. Zhou, M. Zhu, M. Wuttig, Atom Probe Tomography: a Local Probe for Chemical Bonds in Solids, Adv. Mater. 36 (2024) 2403046. https://doi.org/10.1002/adma.202403046.

[63]    D.W. Saxey, Correlated ion analysis and the interpretation of atom probe mass spectra, Ultramicroscopy 111 (2011) 473–479. https://doi.org/10.1016/j.ultramic.2010.11.021.

[64]    S. Torkornoo, M. Bohner, I. McCarroll, B. Gault, Optimization of Parameters for Atom Probe Tomography Analysis of β-Tricalcium Phosphates, Microsc. Microanal. 30 (2024) 1074–1082. https://doi.org/10.1093/mam/ozae077.

[65]    B. Gault, D.W. Saxey, M.W. Ashton, S.B. Sinnott, A.N. Chiaramonti, M.P. Moody, D.K. Schreiber, Behavior of molecules and molecular ions near a field emitter*, New J. Phys. 18 (2016) 033031. https://doi.org/10.1088/1367-2630/18/3/033031.

[66]    M. Bachhav, F. Danoix, B. Hannoyer, J.M. Bassat, R. Danoix, Investigation of O-18 enriched hematite (α-Fe2O3) by laser assisted atom probe tomography, Int. J. Mass Spectrom. 335 (2013) 57–60. https://doi.org/10.1016/j.ijms.2012.10.012.

[67]    T. Kinno, M. Tomita, T. Ohkubo, S. Takeno, K. Hono, Laser-assisted atom probe tomography of 18O-enriched oxide thin film for quantitative analysis of oxygen, Appl. Surf. Sci. 290 (2014) 194–198. https://doi.org/10.1016/j.apsusc.2013.11.039.

[68]    M.P. Moody, L.T. Stephenson, A.V. Ceguerra, S.P. Ringer, Quantitative binomial distribution analyses of nanoscale like-solute atom clustering and segregation in atom probe tomography data, Microsc. Res. Tech. 71 (2008) 542–550. https://doi.org/10.1002/jemt.20582.

[69]   R. Kirchhofer, M.C. Teague, B.P. Gorman, Thermal effects on mass and spatial resolution during laser pulse atom probe tomography of cerium oxide, J. Nucl. Mater. 436 (2013) 23–28. https://doi.org/10.1016/j.jnucmat.2012.12.052.

[70]   D.K. Schreiber, A.N. Chiaramonti, L.M. Gordon, K. Kruska, Applicability of post-ionization theory to laser-assisted field evaporation of magnetite, Appl. Phys. Lett. 105 (2014) 244106. https://doi.org/10.1063/1.4904802.

[71]   B. Gault, A. Chiaramonti, O. Cojocaru-Mirédin, P. Stender, R. Dubosq, C. Freysoldt, S.K. Makineni, T. Li, M. Moody, J.M. Cairney, Atom probe tomography, Nat. Rev. Methods Primer 1 (2021) 51. https://doi.org/10.1038/s43586-021-00047-w.

[72]   M. Herbig, Atomic-Scale Quantification of Grain Boundary Segregation in Nanocrystalline Material, Phys. Rev. Lett. 112 (2014). https://doi.org/10.1103/PhysRevLett.112.126103.

[73]   X. Wang, C. Hatzoglou, B. Sneed, Z. Fan, W. Guo, K. Jin, D. Chen, H. Bei, Y. Wang, W.J. Weber, Y. Zhang, B. Gault, K.L. More, F. Vurpillot, J.D. Poplawsky, Interpreting nanovoids in atom probe tomography data for accurate local compositional measurements, Nat. Commun. 11 (2020) 1022. https://doi.org/10.1038/s41467-020-14832-w.

[74]   M.K. Miller, M.G. Hetherington, Local magnification effects in the atom probe, Surf. Sci. 246 (1991) 442–449. https://doi.org/10.1016/0039-6028(91)90449-3.

[75]   F. Vurpillot, A. Bostel, D. Blavette, Trajectory overlaps and local magnification in three-dimensional atom probe, Appl. Phys. Lett. 76 (2000) 3127–3129. https://doi.org/10.1063/1.126545.

[76]   G. Sha, A. Cerezo, Field ion microscopy and 3-D atom probe analysis of Al3Zr particles in 7050 Al alloy, Ultramicroscopy 102 (2005) 151–159. https://doi.org/10.1016/j.ultramic.2004.09.006.

[77]   R. Lawitzki, P. Stender, G. Schmitz, Compensating Local Magnifications in Atom Probe Tomography for Accurate Analysis of Nano-Sized Precipitates, Microsc. Microanal. 27 (2021) 499–510. https://doi.org/10.1017/S1431927621000180.

[78]   Y.J. Li, A. Kostka, A. Savan, A. Ludwig, Phase decomposition in a nanocrystalline CrCoNi alloy, Scr. Mater. 188 (2020) 259–263. https://doi.org/10.1016/j.scriptamat.2020.07.054.

[79]   S. Ndiaye, C. Bacchi, B. Klaes, M. Canino, F. Vurpillot, L. Rigutti, Surface Dynamics of Field Evaporation in Silicon Carbide, J. Phys. Chem. C 127 (2023) 5467–5478. https://doi.org/10.1021/acs.jpcc.2c08908.

[80]   F. De Geuser, B. Gault, A. Bostel, F. Vurpillot, Correlated field evaporation as seen by atom probe tomography, Surf. Sci. 601 (2007) 536–543. https://doi.org/10.1016/j.susc.2006.10.019.

[81]   B. Gault, M. Müller, A. La Fontaine, M.P. Moody, A. Shariq, A. Cerezo, S.P. Ringer, G.D.W. Smith, Influence of surface migration on the spatial resolution of pulsed laser atom probe tomography, J. Appl. Phys. 108 (2010) 044904. https://doi.org/10.1063/1.3462399.

[82]   B. Gault, F. Danoix, K. Hoummada, D. Mangelinck, H. Leitner, Impact of directional walk on atom probe microanalysis, Ultramicroscopy 113 (2012) 182–191. https://doi.org/10.1016/j.ultramic.2011.06.005.

[83]   C. Oberdorfer, P. Stender, C. Reinke, G. Schmitz, Laser-Assisted Atom Probe Tomography of Oxide Materials, Microsc. Microanal. 13 (2007) 342–346. https://doi.org/10.1017/S1431927607070274.

[84]   Z. Peng, F. Vurpillot, P.-P. Choi, Y. Li, D. Raabe, B. Gault, On the detection of multiple events in atom probe tomography, Ultramicroscopy 189 (2018) 54–60. https://doi.org/10.1016/j.ultramic.2018.03.018.

[85]   S. Korneychuk, C. Grosselindemann, N.H. Menzler, A. Weber, A. Pundt, In-situ TEM reduction of a solid oxide cell with NiO/YSZ fuel electrode, J. Power Sources 625 (2025) 235626. https://doi.org/10.1016/j.jpowsour.2024.235626.

[86]   J.A. Rodriguez, J.C. Hanson, A.I. Frenkel, J.Y. Kim, M. Pérez, Experimental and Theoretical Studies on the Reaction of $H_2$ with NiO: Role of O Vacancies and Mechanism for Oxide Reduction, J. Am. Chem. Soc. 124 (2002) 346–354. https://doi.org/10.1021/ja0121080.

[87]   B. Li, Y. Wei, H. Wang, Non-isothermal reduction kinetics of $Fe_2O_3$–NiO composites for formation of Fe–Ni alloy using carbon monoxide, Trans. Nonferrous Met. Soc. China 24 (2014) 3710–3715. https://doi.org/10.1016/S1003-6326(14)63519-6.

[88]   I. Dirba, C.A. Schwöbel, A. Zintler, P. Komissinskiy, L. Molina-Luna, O. Gutfleisch, Production of Fe nanoparticles from γ-$Fe_2O_3$ by high-pressure hydrogen reduction, Nanoscale Adv. 2 (2020) 4777–4784. https://doi.org/10.1039/D0NA00635A.

[89]   S. Pöyhtäri, J. Ruokoja, E.-P. Heikkinen, A. Heikkilä, T. Kokkonen, P. Tynjälä, Kinetic Analysis of Hydrogen Reduction of Nickel Compounds, Metall. Mater. Trans. B 55 (2024) 251–265. https://doi.org/10.1007/s11663-023-02955-6.

[90]   D. Yu, M. Zhu, T.A. Utigard, M. Barati, TGA kinetic study on the hydrogen reduction of an iron nickel oxide, Miner. Eng. 54 (2013) 32–38. https://doi.org/10.1016/j.mineng.2013.03.018.

[91]  K.V. Manukyan, A.G. Avetisyan, C.E. Shuck, H.A. Chatilyan, S. Rouvimov, S.L. Kharatyan, A.S. Mukasyan, Nickel Oxide Reduction by Hydrogen: Kinetics and Structural Transformations, J. Phys. Chem. C 119 (2015) 16131–16138. https://doi.org/10.1021/acs.jpcc.5b04313.

[92]  Q. Jeangros, T.W. Hansen, J.B. Wagner, C.D. Damsgaard, R.E. Dunin-Borkowski, C. Hébert, J. Van herle, A. Hessler-Wyser, Reduction of nickel oxide particles by hydrogen studied in an environmental TEM, J. Mater. Sci. 48 (2013) 2893–2907. https://doi.org/10.1007/s10853-012-7001-2.

[93]  G.H. La, S.H. Lee, D.J. Min, Fundamental study on preferential reduction of Ni in NiFe2O4, Miner. Eng. 164 (2021) 106829. https://doi.org/10.1016/j.mineng.2021.106829.

[94]  D. Spreitzer, J. Schenk, Reduction of Iron Oxides with Hydrogen—A Review, Steel Res. Int. 90 (2019) 1900108. https://doi.org/10.1002/srin.201900108.

[95]  G.I. Silman, Compilative Fe – Ni phase diagram with author's correction, Met. Sci. Heat Treat. 54 (2012) 105–112. https://doi.org/10.1007/s11041-012-9463-x.

[96]  K. Takanashi, M. Mizuguchi, T. Kojima, T. Tashiro, Fabrication and characterization of L10-ordered FeNi thin films, J. Phys. Appl. Phys. 50 (2017) 483002. https://doi.org/10.1088/1361-6463/aa8ff6.

[97]  E. Lima, V. Drago, P.F.P. Fichtner, P.H.P. Domingues, Tetrataenite and other Fe–Ni equilibrium phases produced by reduction of nanocrystalline NiFe2O4, Solid State Commun. 128 (2003) 345–350. https://doi.org/10.1016/j.ssc.2003.08.046.

[98]  G. Varvaro, P. Imperatori, S. Laureti, D. Peddis, F. Locardi, M. Ferretti, C. Cannas, M.S. Angotzi, N. Yaacoub, A. Capobianchi, Facile and fast synthesis of highly ordered L10-FeNi nanoparticles, Scr. Mater. 238 (2024) 115754. https://doi.org/10.1016/j.scriptamat.2023.115754.

[99]  S.V. Divinski, Diffusion in Nanostructured Materials, Defect Diffus. Forum 289–292 (2009) 623–632. https://doi.org/10.4028/www.scientific.net/DDF.289-292.623.

[100]  T.J. Chuang, K. Wandelt, Study of interdiffusion of Ni/Fe layers by Auger Sputter profiling, Surf. Sci. 81 (1979) 355–369. https://doi.org/10.1016/0039-6028(79)90105-5.

[101]  J. Liu, K. Barmak, Interdiffusion in nanometric Fe/Ni multilayer films, J. Vac. Sci. Technol. A 33 (2015) 021510. https://doi.org/10.1116/1.4905465.

[102]  R.S. Singh, S. Guruswamy, Deformation behavior and magnetic properties of equiatomic FeNi single crystals, AIP Adv. 14 (2024) 045206. https://doi.org/10.1063/5.0196977.

[103] Y.J. Li, P. Choi, S. Goto, C. Borchers, D. Raabe, R. Kirchheim, Evolution of strength and microstructure during annealing of heavily cold-drawn 6.3 GPa hypereutectoid pearlitic steel wire, Acta Mater. 60 (2012) 4005–4016. https://doi.org/10.1016/j.actamat.2012.03.006.

**Supplementary materials**

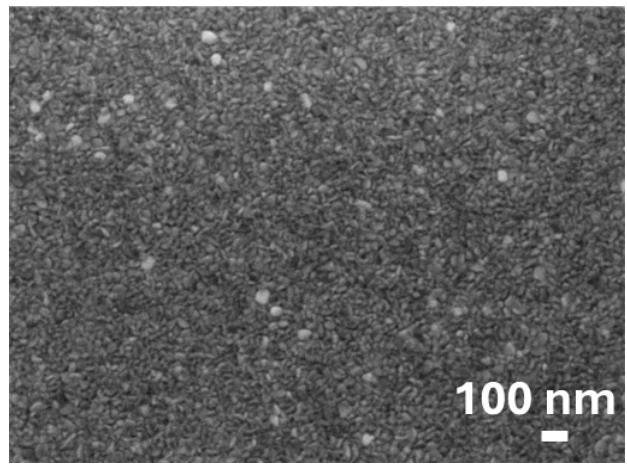

*Figure S1*. High resolution SE image showing the nanoscale grains of the as-deposited $Fe_{50}Ni_{50}O_x$ thin film.

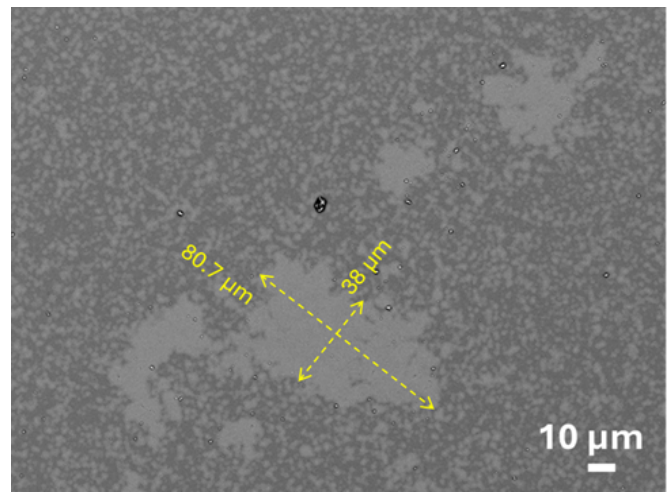

*Figure S2*. BSE image of the $Fe_{50}Ni_{50}O_x$ thin film after reduction at 280°C for 7 min showing significant coarsening of the reduction product phase into large islands in some parts of the microstructure.

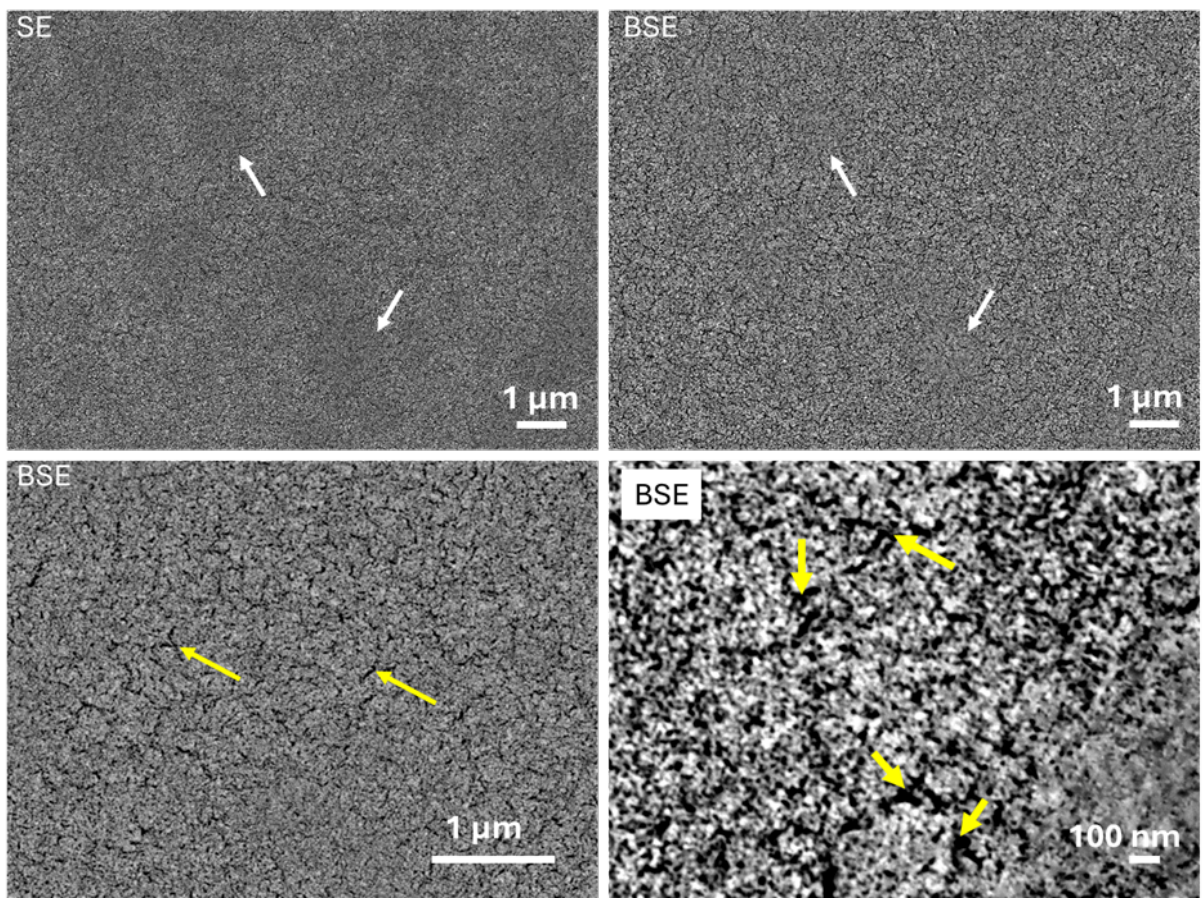

**Figure S3**. SEM images of the thin film after 10 min reduction at 280°C at different magnifications showing extensive reduction resulting in a highly porous (yellow arrows) microstructure, with pockets of different contrast (some highlighted by white arrows), presumably corresponding to partially reduced oxide regions.

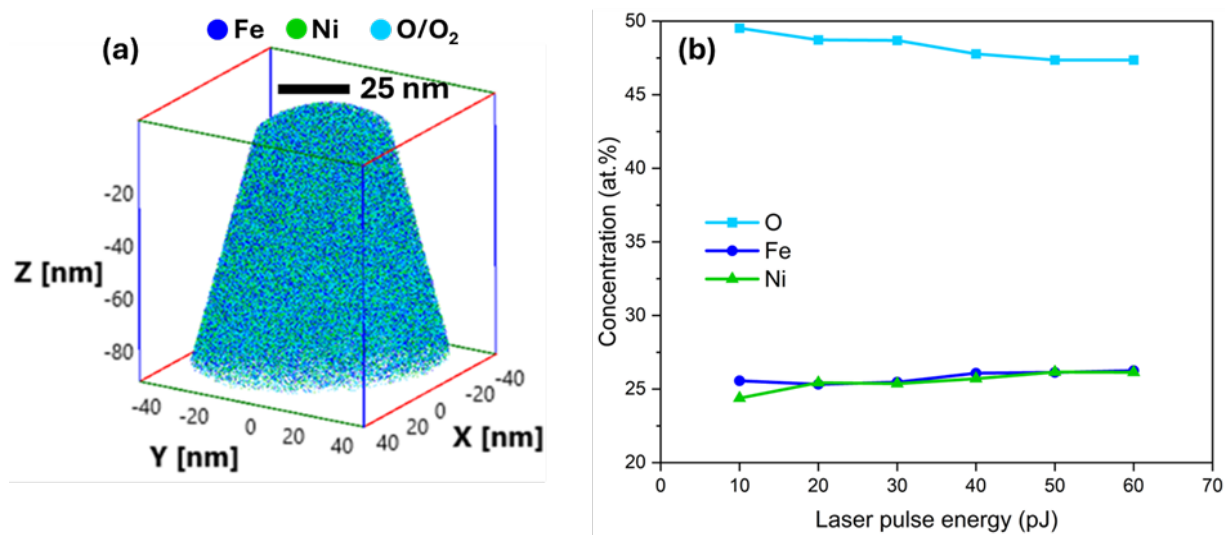

**Figure S4**. APT analysis of the as-deposited thin film: a) 3D reconstruction of the data collected at a laser energy and pulse frequency of 50 pJ and 125 kHz respectively. b) Graph showing the variation of elemental composition with LPE due to O loss through molecular ion dissociation into neutral oxygen. The data were collected with a pulse frequency of 200 kHz, apart from that at 50 pJ which was collected with a frequency of 125 kHz.

*Table S1. Bulk composition of the as-deposited film analysed using a LPE of 50 pJ and a pulse frequency of 125 kHz. Note that Ga is from ion implantation during FIB lift-out and H ions are often present in APT datasets and comes from various sources including adsorbed H on the surface of the specimen or some residual $H_2$ inside the ultra-high vacuum atom probe's analysis chamber.*

| Atom type | Count | Concentration (at.%) |
| --- | --- | --- |
| Fe | 2,018,761.80 | 25.75 |
| Ni | 2,020,477.00 | 25.77 |
| O | 3,665,129.30 | 46.74 |
| Cu | 12,179.70 | 0.16 |
| H | 107,240.60 | 1.37 |
| Ga | 16,550.60 | 0.21 |

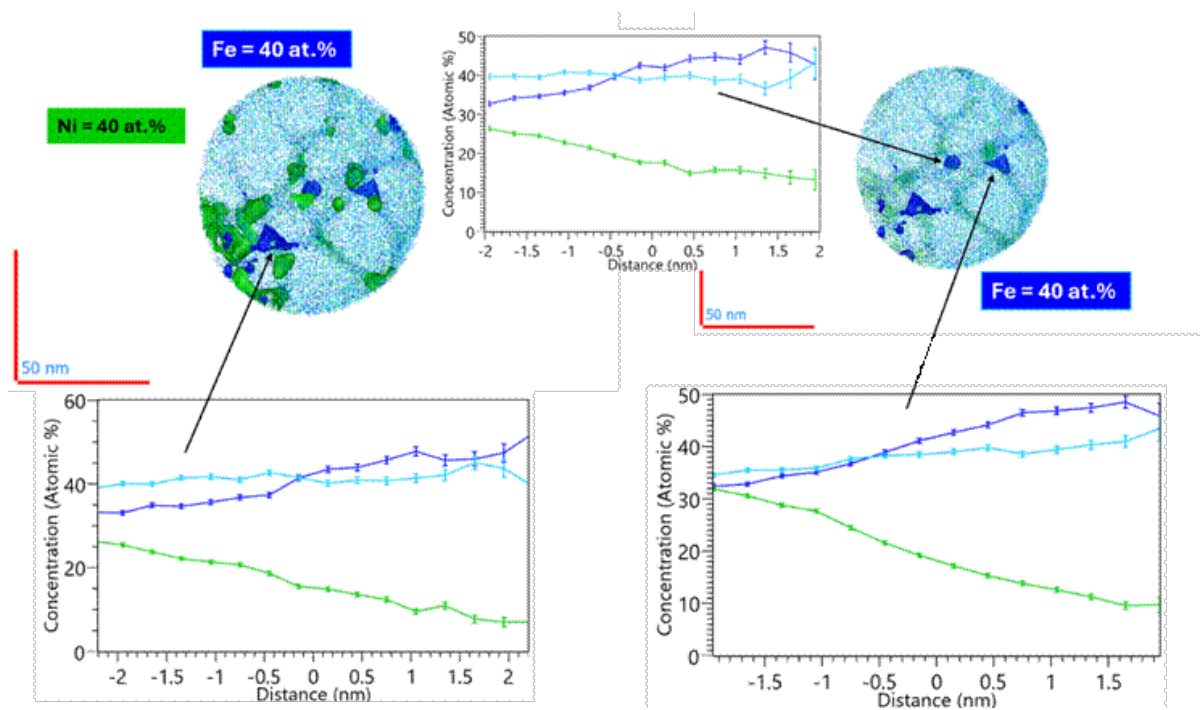

*Figure S5. Proximity histograms of different Fe (40 at.%) isosurfaces.*

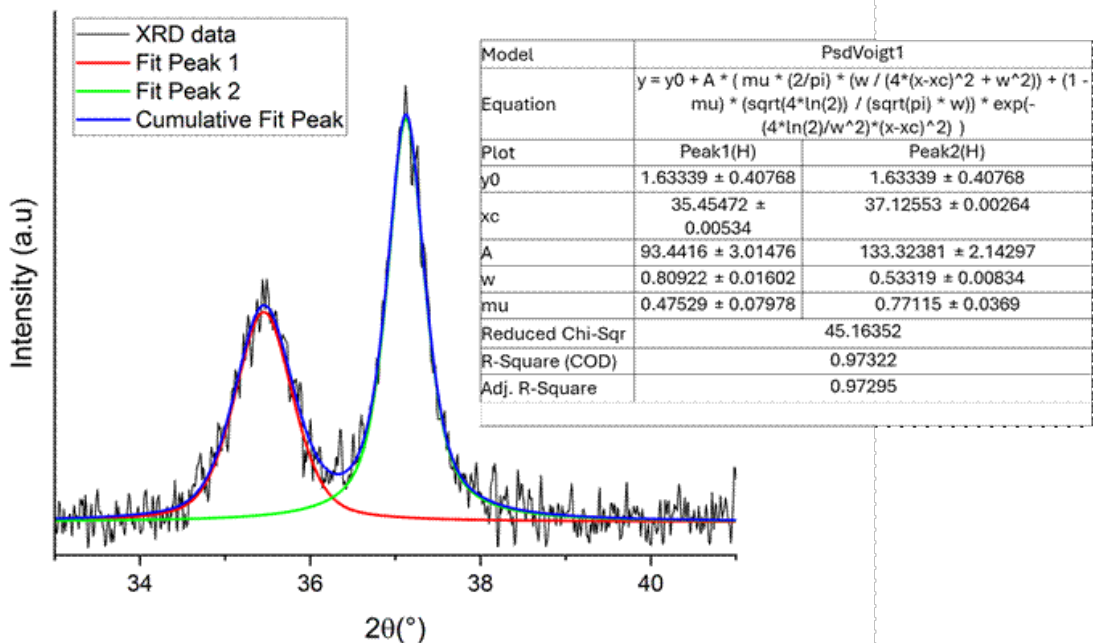

*Figure S6. Some of the XRD peaks were fitted with a Pseudo-Voigt function for FWHM estimation later used in the Scherrer's equation for the estimation of the crystallite size.*

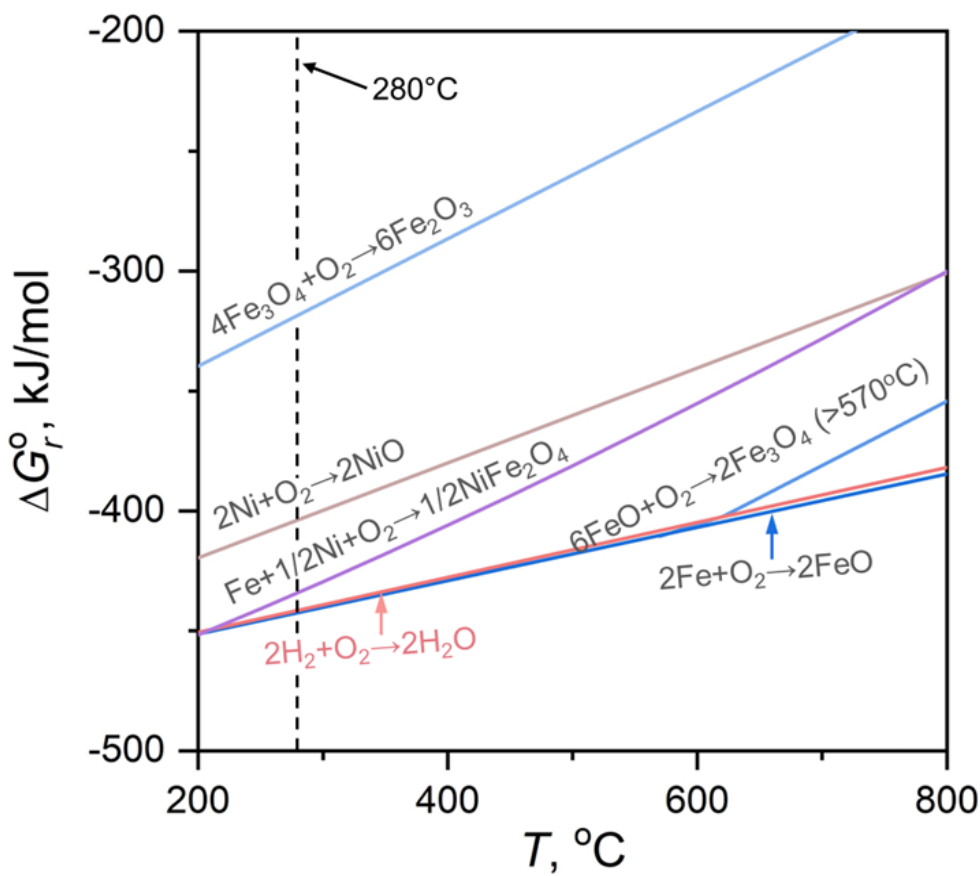

*Figure S7. Ellingham-Richardson diagram for NiO, NiFe$_2$O$_4$, and various Fe oxides (Fe$_2$O$_3$, Fe$_3$O$_4$, and FeO) under 1 atm. Note that the diagram was generated using Thermo-Calc software from thermodynamic data of bulk oxides.*